\documentclass[aip,jap,numerical,reprint]{revtex4-1}
\pdfoutput=1  
\usepackage{dcolumn}
\usepackage{graphicx,psfrag,epsf,epsfig}
\usepackage{bm,bbm}
\usepackage{latexsym}
\usepackage{mathrsfs,textcomp}
\usepackage{amsmath,amssymb}

\usepackage{float}
\usepackage{hyperref}

\begin{document}


\title[]{Impact of nanostructure configuration on the photovoltaic performance of quantum dot arrays}

\author{Aude Berbezier}
\author{Urs Aeberhard*}
 \email{u.aeberhard@fz-juelich.de}
\affiliation{
IEK5-Photovoltaik, Forschungszentrum J\"ulich, 52425 J\"ulich, Germany
}%



\begin{abstract}

In this work, a mesoscopic model based on the
non-equilibrium Green's function formalism for a tight-binding-like effective Hamiltonian is used to investigate a selectively
contacted quantum dot array designed for operation as a single junction quantum dot solar cell. By establishing a direct relation between nanostructure configuration and optoelectronic properties, the investigation reveals the influence of 
inter-dot and dot-contact coupling strengths on the rates of charge carrier photogeneration, radiative recombination, and extraction at contacts, and
consequently on the ultimate performance of photovoltaic devices
with finite quantum dot arrays as the active medium. For long carrier lifetimes, the dominant configuration effects originate in the dependence of the joint density of states on the inter-dot coupling in terms of band width and effective band gap. In the low carrier lifetime regime, where recombination competes with carrier extraction, the extraction efficiency shows a critical dependence on the dot-contact coupling.
\end{abstract}

\maketitle

\noindent

\section{Introduction}

In a large variety of concepts for next generation solar cells, such as devices based on intermediate bands, hot carriers or multiple exciton generation, nanostructures like quantum wells, wires or dots provide the
specific physical properties required to reach beyond current
efficiency limitations\cite{green:00,marti:04,tsakalakos:08}. Among these nanostructures, quantum dots (QD) are of special interest due to their largely tunable optoelectronic characteristics, and regimented QD structures thus represent promising candidates for the
implementation of novel high efficiency solar cell concepts\cite{Aroutiounian2001,nozik:02,Raffaelle2002,marti:06_tsf,Oshima2008a,Conibeer2008}. 

However, to enable exploitation of the design degrees of freedom provided, the complex relation between structural configuration parameters and device
characteristics requires thorough investigation. In conventional
detailed-balance modeling approaches, this connection is provided only indirectly via the local variation of macroscopic bulk material parameters
such as absorption coefficient, mobility and lifetime of charge
carriers\cite{Aroutiounian2005,Shao2007,Gioannini2013}. For a direct, consistent and comprehensive assessment of
configuration-related photovoltaic performance, both inter-band
dynamics and transport of charge carriers need to be described in a
unified microscopic picture, like the one provided by the
non-equilibrium Green's function (NEGF) formalism\cite{ae:jcel_11}.
Since a fully microscopic treatment of extended systems of coupled
QD comes with a huge computational cost, application of multi-scale
approaches is required, such as the mesoscopic QD orbital 
tight-binding model introduced in Ref.~\onlinecite{ae:oqel_12}.

\begin{figure}[t!]
\begin{center}
\includegraphics[width=0.4\textwidth]{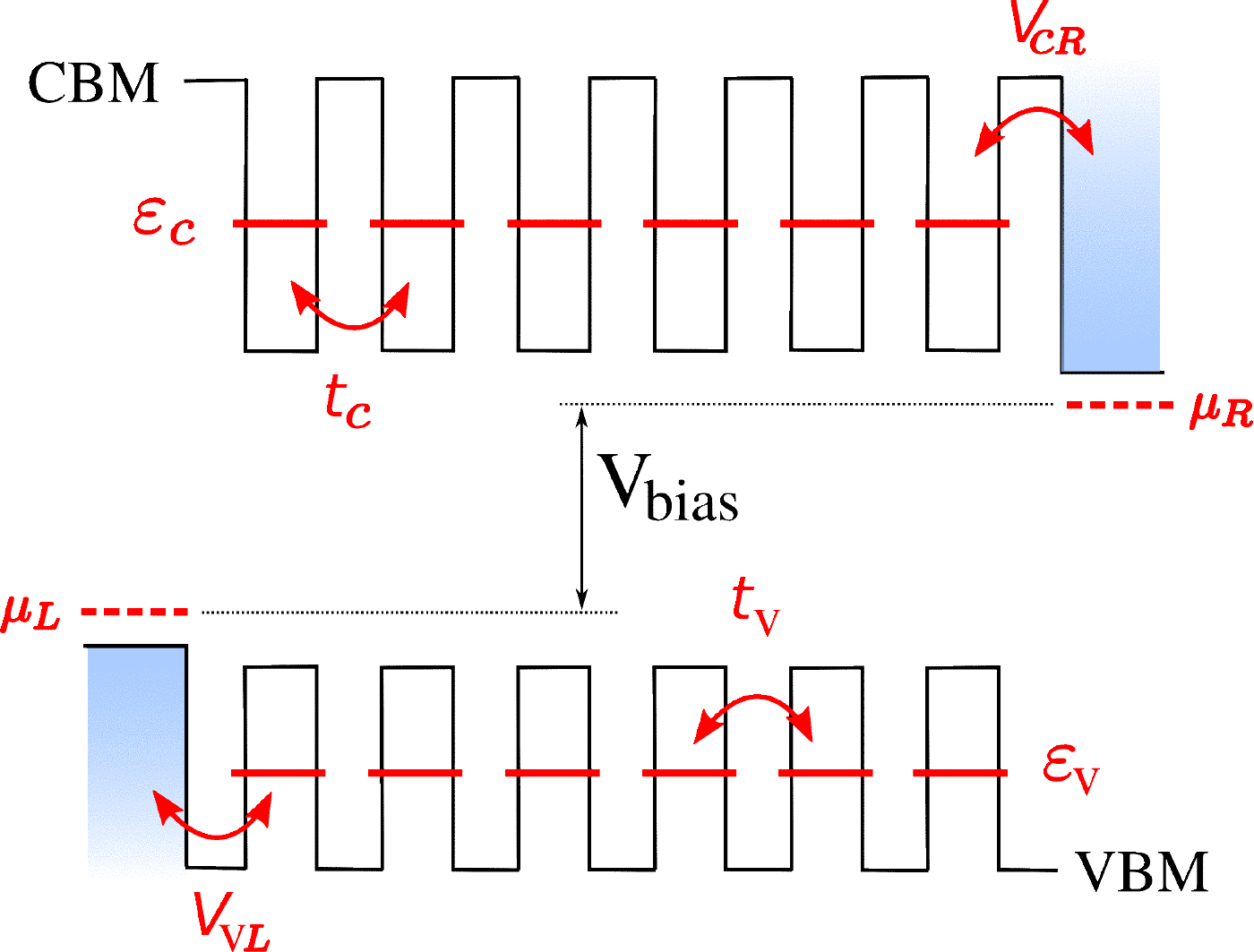}
\caption{(color online) Schematic band diagram and phenomenological parameters
characterizing the selectively contacted QD array. Isolated QD states at energies $\varepsilon_{c,v}$ are connected through inter-dot coupling terms $t_{c,v}$. To enable photocarrier extraction and injection of dark current under applied bias voltage $V_{\textrm{bias}}$, the QD states are connected to contacts via coupling elements $V_{B}$. The carrier
selectivity of the contacts, i.e., $V_{cL}=V_{vR}=0$, is  required for photocurrent rectification at flat band conditions in the absence of a doping
induced junction.  While left and right contacts are assumed to be
equilibrated with charge carrier populations characterized by
chemical potentials $\mu_{L}$ and $\mu_{R}=\mu_{L}+V_{\textrm{bias}}$, respectively, the
carrier population inside the absorber emerges entirely from the
quantum-statistical formalism applied. \label{fig:parameters}}
\end{center}
\end{figure}

Here, this model is used in a simplified version, with restriction
to a single basis orbital per band, phenomenological nearest-neighbor
hopping parameters and flat band conditions. These simplifications
assure maximum transparency in the analysis of the photovoltaic
mechanisms in the QD arrays considered and allow for direct comparison with the conventionally used periodic picture of an infinite QD chain. In this way, clear relations can be established
between the radiative limit of photovoltaic device performance and configuration dependent properties such as the finite
system size in terms of a limited number of QD layers, the coupling between QDs, and the presence and nature of
contacts.

The article is organized as follows. In Sec.~\ref{sec:model}, the model for the electronic structure and the dynamics of charge carriers in selectively contacted quantum dot arrays under illumination is discussed in detail. The focus is on properties relevant for photovoltaic device operation, such as absorption cross section, emission spectra, and carrier extraction efficiency. In Sec.~\ref{sec:results}, the numerical implementation of the model is evaluated for a range of inter-dot and dot-contact coupling values for both long and short charge carrier lifetimes. The results are briefly summarized in a last section.

\section{Model\label{sec:model}}

\subsection{Hamiltonian and NEGF equations}

The phenomenological Hamiltonian for
the device sketched in Fig.~\ref{fig:parameters} reads
($b\in\{c,v\}$):
\begin{align}
\hat{\mathcal{H}}_{0}^{b}=&\sum_{i=1}^{N_{QD}-1}
t_{b,ii+1}\Big[\hat{d}_{b,i+1}^{\dagger}\hat{d}_{b,i}+h.c.\Big]+\sum_{i=1}^{N_{QD}}\varepsilon_{b,i}\hat{n}_{b,i}\label{eq:ham}\\
&\equiv \sum_{i,j=1}^{N_{QD}}H^b_{0,ij}\hat{d}_{b,i}^{\dagger}\hat{d}_{b,j},
\end{align}
where $N_{QD}$ is the number of dots, $t_{b}$ is the inter-dot coupling, $\hat{n}_{b}\equiv \hat{d}_{b}^{\dagger}\hat{d}_{b}$ is the carrier density operator
and $\varepsilon_{b}$ is the QD energy level. The phenomenological parameters can, in principle, be determined from a microscopic theory of the QD states for a given nanostructure configuration, e.g., via QD Wannier functions \cite{vukmirovic:07}, but this is beyond the scope of the present paper. Here, we will concentrate on the impact of the magnitude of these parameters on the optoelectronic properties that are relevant for the photovoltaic performance. An estimate for a realistic range of inter-dot coupling strengths can be obtained, e.g., from the coefficients of a cosine Fourier expansion of the dispersion relation for superlattice Bloch states \cite{wacker:02}. 

The Hamiltonian in \eqref{eq:ham} is used in the equations for the steady state
charge carrier NEGF, which read
for a given band and energy $E$:
\begin{align}
\mathbf{G}^{R}(E)&=\left[\big\{\mathbf{G}_{0}^{R}(E)\big\}^{-1}
-\mathbf{\Sigma}^{R I}(E)-
\mathbf{\Sigma}^{RB}(E)\right]^{-1},\label{eq:retgf}\\
\mathbf{G}^{\lessgtr}(E)&=\mathbf{G}^{R}(E)
\left[\mathbf{\Sigma}^{\lessgtr I}(E)
+\mathbf{\Sigma}^{\lessgtr
B}(E)\right]\mathbf{G}^{A}(E),\label{eq:corrf}
\end{align}
where 
\begin{align}
\mathbf{G}_{0}^{R}(E)=\left[(E+i\eta)\mathbbm{1}-\mathbf{H}_{0}\right]^{-1},~\mathbf{G}^{A}=[\mathbf{G}^{R}]^{\dagger} 
\end{align} 
and $\mathbf{G}\equiv\left[G_{ij}\right]$ with
\begin{align}
G_{ij}(E)=&\int d\tau e^{iE\tau/\hbar}G_{ij}(\tau),\quad \tau=t'-t
\end{align}
and
\begin{align}
G_{ij}(t,t')=-\frac{i}{\hbar}\langle \hat{T}_{C}\{\hat{d}_{i}(t)\hat{d}_{j}^{\dagger}(t')\}\rangle
\end{align}
is the ensemble average ordered on the Keldysh contour\cite{keldysh:65}.

The self-energy (SE) terms $\Sigma^{\cdot B}$
encode the hybridization of QD and contact states, enabling the description
of charge carrier extraction and injection at the electrodes: 
\begin{align}
\Sigma^{R B}_{ij}(E)=&\delta_{ij}\Big[\Delta_{B}(E)-\frac{i}{2}\Gamma_{B}(E)\Big],\\
\Sigma^{\lessgtr B}_{ij}(E)=&i\left[f(E-\mu_{B})-\frac{1}{2}\pm\frac{1}{2}\right]\delta_{ij}\Gamma(E),\label{eq:contactse}
\end{align}
where $B\in\{L,R\}$, $f$ is the Fermi-Dirac distribution function and
$\mu_{B}$ is the chemical potential of the contact considered. 
The real part $\Delta_{B}$ describes the energetic shift of the electronic states due to coupling  to the electrodes, and the broadening function $\Gamma_{B}$ is related to the coupling
parameter $V_{B}$ and to the local density of states (LDOS) $\rho_{B}$ of the electrode contact layer through $\Gamma_{B}(E)=2\pi V_{B}^{2}\rho_{B}(E)$. In this work, the semi-infinite 1D tight-binding chain description of a finite band-width electrode is used\cite{frederiksen:04}, with:
\begin{align}
\Delta_{B}(E)=&\frac{\Gamma_{0}}{2}\times\left\{\begin{array}{rl}
x,&\quad|x|\leq 1\\
\big(x-\mathrm{sgn}(x)\sqrt{x^2-1}\big),&\quad|x|> 1
\end{array}\right. \\
\Gamma_{B}(E)=&\Gamma_{0}\theta(1-|x|)\sqrt{1-x^2},\\
 \Gamma_{0}=&2V_{B}^2/|t_{B}|,\quad x=(E-\varepsilon_{B})/(2|t_{B}|),
\end{align}
where $\varepsilon_{B}$ is the on-site energy and $t_{B}$ the hopping element of the electrode material. In order to enable efficient charge carrier injection and extraction, overlap of electrode and QD DOS demands $\varepsilon_{B}'\equiv \varepsilon_{B}-2|t_{B}|<\varepsilon_{b}-2|t_{b}|$ (electron case). 

For the numerical example we chose $\{\varepsilon_{v},\varepsilon_{c}\}=\{-0.7,0.6\}$ eV, similar to the values obtained for the lowest states of PbSe colloidal QDs of $\sim 3$ nm diameter using $\mathbf{k}\cdot \mathbf{p}$ band structure calculations\cite{ae:pccp_12,bartnik:07,vaxenburg:12}. The determination of the magnitude of the coupling constant $V_{B}$ usually requires a full microscopic treatment of the contact region. An estimate of the relevant range of coupling values can be obtained from the comparison of the LDOS and density from full solution of the Schr\"odinger equation in this region with the same quantities as obtained by the phenomenological approach above for different values of $V_{B}$.   
In order to achieve the carrier selectivity required for charge
separation and current rectification in the absence of a built-in
potential, the contact couplings are set to zero at the minority
carrier boundaries, which are chosen at the left electrode for
electrons and at the right electrode for holes, i.e., $V_{cL}=V_{vR}=0$ (see also Fig.~\ref{fig:parameters}). For the electrode states, $\{\varepsilon_{vL},\varepsilon_{cR}\}=\{-0.6,0.5\}$ and $|t_{B}|=\hbar^2/(2m_{B}^{*}\Delta^2)$ for $B\in\{vL,cR\}$, with $m_{B}^{*}=0.25~m_{0}$ and $\Delta=2.7$ \AA~are used.

\subsection{Density of states}

\begin{figure}[!t]
\begin{center}
\includegraphics[width=0.48\textwidth]{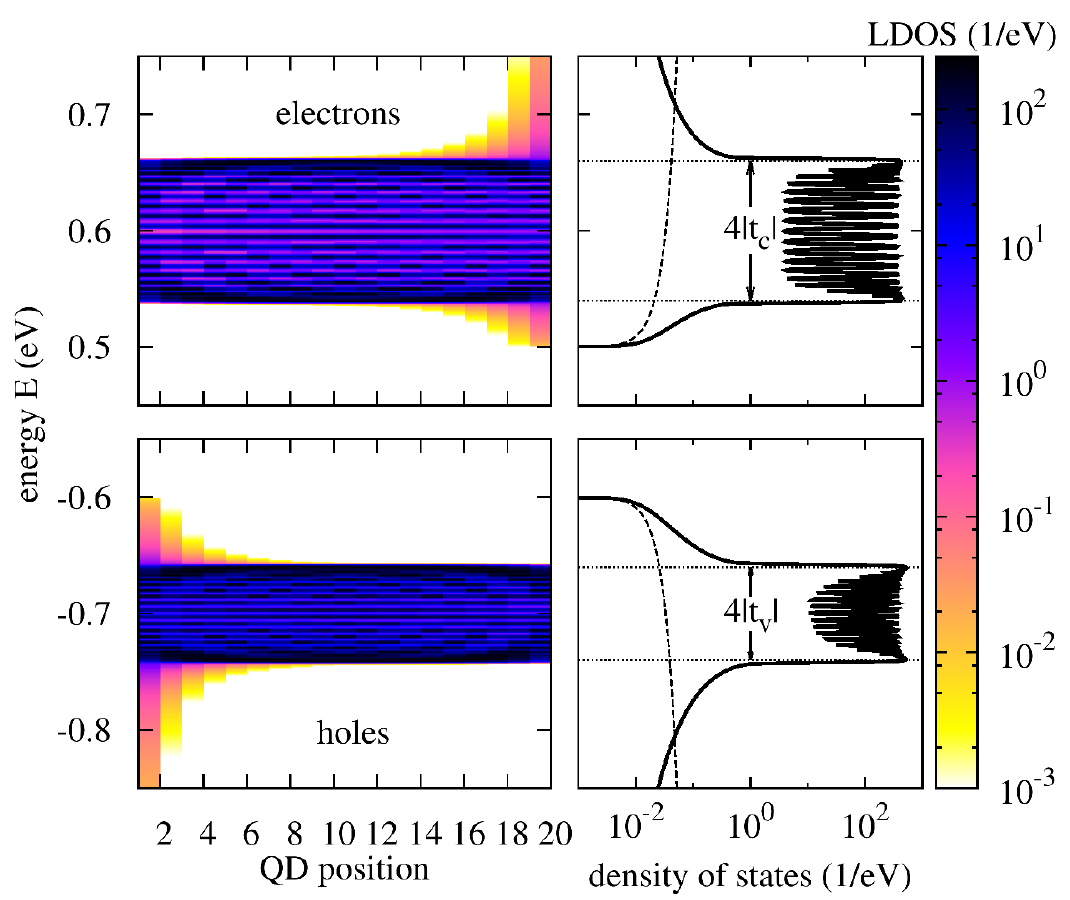}
\caption{(color online) Local (left) and total (right) density of electron and hole states for a selectively
contacted 20 QD array with $V_{cR}=V_{vL}=0.1$ eV, $t_{c}=-0.03$ eV and
$t_{v}=-0.02$ eV. At the closed contacts, the band width approaches the
infinite single orbital tight-binding chain value of $4|t_{b}|$, $b=c,v$ (dotted lines). On the contrary, at
the open contacts, hybridization with contact
states induces a pronounced widening of the bands and a shift of the band edge towards the electrode edges at $\varepsilon'_{cR}=0.5$ eV and $\varepsilon'_{vL}=-0.6$ eV, according to the local density of contact states at the surface of the semi-infinite electrode (dashed lines). 
\label{fig:ldos_2b_eh}}
\end{center}
\end{figure}

\begin{figure}[t]
\begin{center}
\includegraphics[width=0.4\textwidth]{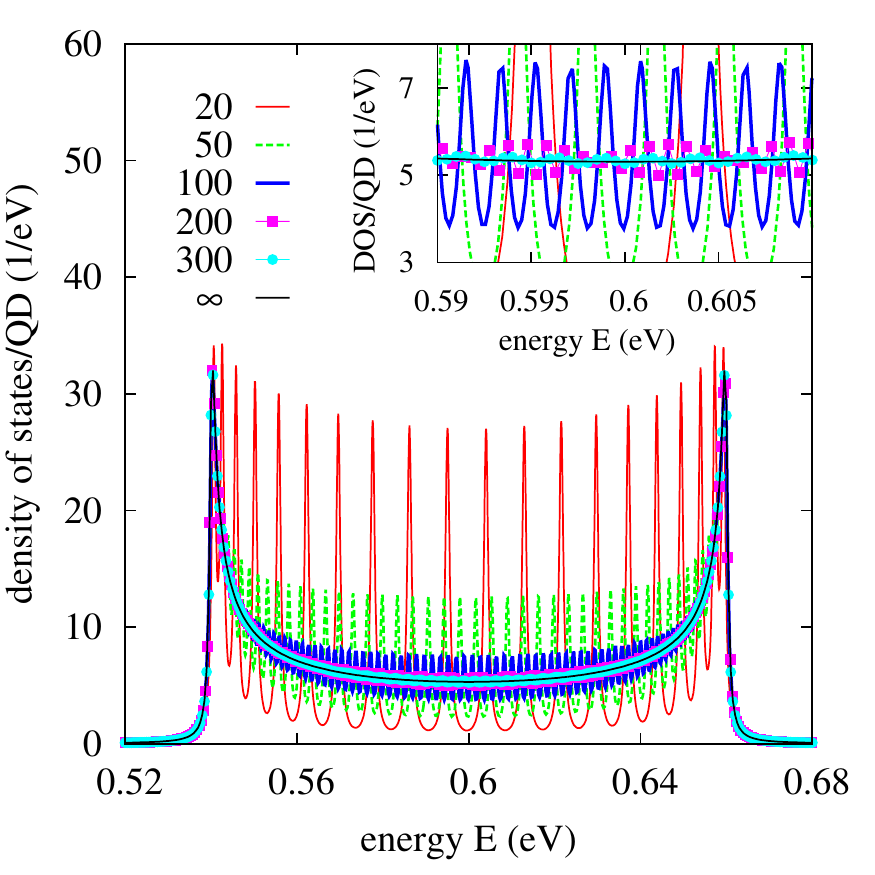}
\caption{(color online) Convergence of the average density of electronic states per QD to the infinite 1D TB chain limit with increasing number of QDs between the contacts, assuming symmetric contact coupling  $V_{cL}=V_{cR}=0.1$ eV, $t_{c}=-0.03$ eV and a homogeneous broadening $\eta_{c}=0.5$ meV.
\label{fig:DOS_tc}}
\end{center}
\end{figure}

At this level, the formalism provides the electronic structure of the non-interacting system under arbitrary non-equilibrium conditions, e.g., via the local density of charge carrier states $\mathcal{\rho}_{i}(E)=-\frac{1}{\pi}\Im \mathrm{m}G_{ii}^{R}(E)$, where $i$ labels the dot position. Figure \ref{fig:ldos_2b_eh} displays the LDOS for an array of 20 coupled QDs, together with the total DOS obtained from summing the individual QD contributions. In contrast to standard simulations of QD-based solar cells, where extended QD superlattices are modeled using periodic boundary conditions\cite{jiang:06,tomic:08}, discrete states can still be distinguished in the total DOS. On the other hand, the main effect of the presence of a contact is the broadening and shifting of the LDOS according to the electrode DOS (dashed lines), especially in the vicinity of the electrode, where the band width is strongly increased from the limiting value of $\sim 4|t_{b}|$ ($b=c,v$) of the infinite single orbital tight-binding (TB) chain (dotted horizontal lines). This limiting value is nearly recovered in arrays with as few as 20 QDs. The energy integration of the total DOS yields $N_{QD}$, in agreement with the associated sum rule. The effects of finite system size persist up to a large number of QDs, as can be verified in Fig.~\ref{fig:DOS_tc} displaying the convergence of the average electron DOS per QD with increasing number of QDs towards the infinite 1D TB chain limit given by 
\begin{align}
\bar{\rho}_{c,\infty}(E)=\left[2\pi |t_{c}|\sqrt{1-\left(\frac{E+i\eta_{c}-\varepsilon_{c}}{2|t_{c}|}\right)^2}\right]^{-1},\label{eq:1dtb_dos}
\end{align}
for which a homogeneous broadening of $\eta_{c}=0.5$ meV was assumed. Thus, while the band width of the finite system is almost identical to that of the infinite chain, the spectral distribution of the DOS differs considerably even for moderate QD numbers, from clearly distinguishable contributions of individual dots to a smooth continuum.

\begin{figure}[t!]
\begin{center}
\includegraphics[width=0.4\textwidth]{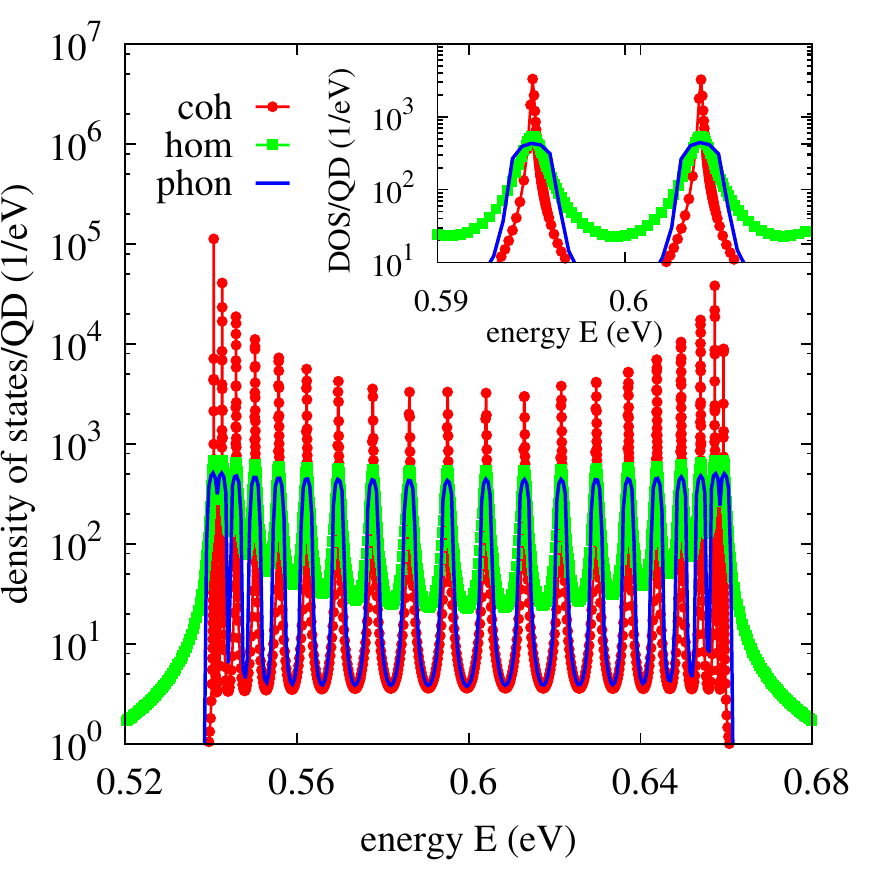}
\caption{(color online) Site-integrated density of electron states for a 20 QD array ($V_{cL}=V_{cR}=0.1$ eV, $t_{c}=-0.03$ eV) with coupling induced broadening only ("coh"), constant homogeneous broadening $\eta=0.5$ meV ("hom") and for quasi-elastic coupling to optical phonons of energy $\hbar\Omega=0.2$ meV ("phon").
\label{fig:DOS_broad}}
\end{center}
\end{figure}

The SE terms $\Sigma^{\cdot I}$ include the
interactions relevant for the photovoltaic device operation, i.e.,
coupling to photons, phonons and to other charge
carriers (not considered in this work). Since photovoltaic devices are operated at or above room temperature, thermal broadening is considered via the physical
line-shape provided by quasi-elastic coupling to phonons. Indeed, at the coherent limit, the weak inter-dot and dot-contact coupling and resulting strong localization of charge carriers on the QDs would lead to extremely sharp resonances in the spectral quantities, exhibiting line widths far below the thermal energy at room temperature. For the computation of static spectral quantities such as the electronic density of states, a homogeneous broadening can be introduced via the damping parameter $\eta$. However, this is not a valid approach for transport simulations, as it violates current conservation\cite{lake:97}. Therefore, for the evaluation of transport properties, broadening induced by scattering of charge carriers and optical phonons is considered on the level of the current conserving self-consistent Born approximation for the coupling of interacting fermions to a non-interacting phonon bath\footnote{A perturbative approach might not always apply in QD array systems, in some cases of strong electron-phonon interaction, a polaronic picture might be more appropriate; in the present work, this description is merely used as a phenomenological, but still current conserving broadening mechanism}. Since at this stage, no real phonon modes are considered, a constant effective coupling matrix element $|\bar{\mathcal{M}}_{ph}|^2\approx1$  meV diagonal in the electronic indices and a phonon frequency $\hbar\Omega\approx 0.2-0.5$ meV is assumed, which provides a quasi-elastic scattering resulting in a local inhomogeneous broadening. The corresponding self-energies have the form:
\begin{align}
\boldsymbol{\Sigma}^{\lessgtr}(E)=|\bar{\mathcal{M}}_{ph}|^2&\Big[N_{ph}\mathbf{G}^{\lessgtr}(E\mp\hbar\Omega)
\nonumber\\&+(N_{ph}+1)\mathbf{G}^{\lessgtr}(E\pm\hbar\Omega)\Big],\\
\mathbf{\Sigma}^{R}(E)=|\bar{\mathcal{M}}_{ph}|^2&\Big
[(N_{ph}+1)\mathbf{G}^{R}(E-\hbar\Omega)\nonumber\\&+N_{ph}\mathbf{G}^{R}(E
+\hbar\Omega)\nonumber\\&+\frac{1}{2}\big\{\mathbf{G}^{<}(E+\hbar\Omega)-
\mathbf{G}^{<}(E-\hbar\Omega)\big\}\Big],
\end{align}
where the phonon mode occupation number is given by the Bose-Einstein function
$N_{ph}=(e^{\hbar\Omega/k_{B}T}-1)^{-1}$. 
Fig.~\ref{fig:DOS_broad} displays the integrated density of states for different broadening mechanisms, i.e., for broadening induced by inter-dot and dot contact coupling only ("coh"), for homogeneous broadening with $\eta=0.5$ meV ("hom") and for quasi-elastic coupling to optical phonons of energy $\hbar\Omega=0.2$ meV ("phon").

\subsection{Absorption and photocurrent generation}

\begin{figure}[t]
\begin{center}
\includegraphics[width=0.4\textwidth]{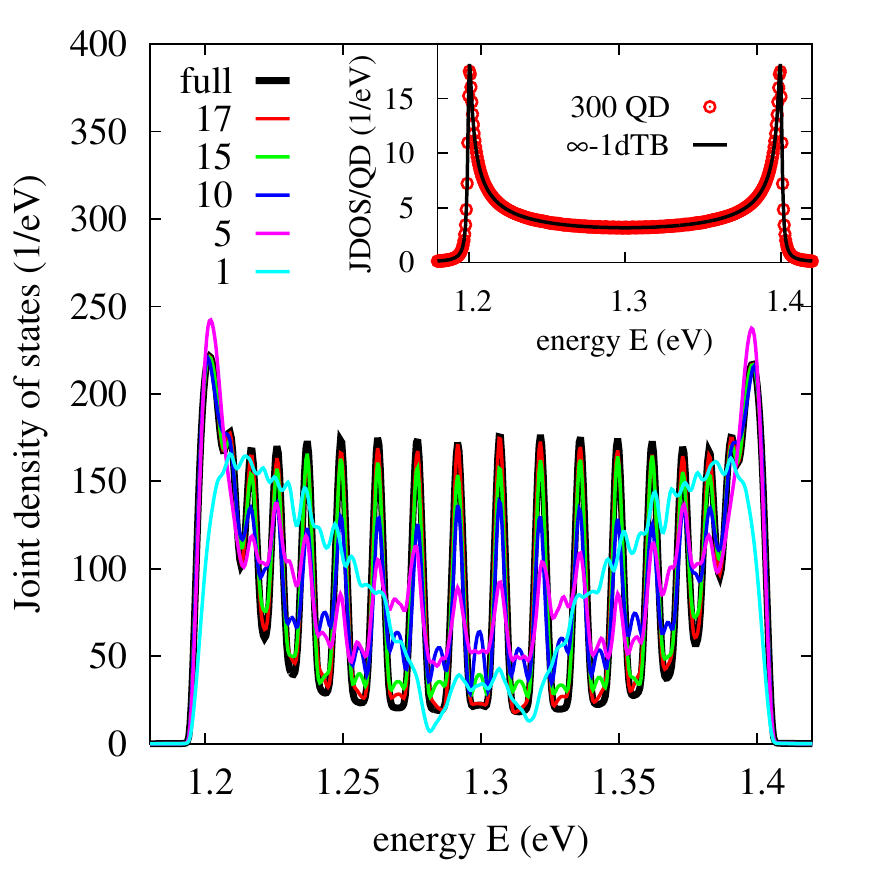}
\caption{(color online) Convergence of the JDOS of a 20 QD system with increasing non-locality and to the infinite 1D TB chain for large system size (inset).
\label{fig:nonloc}}
\end{center}
\end{figure}

In the coupling
of charge carriers to electromagnetic fields, stimulated
and spontaneous processes need to be
distinguished. Photogeneration and stimulated emission induced by incident radiation is described via the self-energy for the interaction with a classical field with vector potential $\mathbf{A}$:\cite{ae:prb89_14}
\begin{align}
&\Sigma^{e\gamma-st,\lessgtr}_{ij,\eta}(E,\hbar\omega)=\mathcal{M}^{e\gamma,\eta}_{i}\mathcal{M}^{e\gamma,\eta}_{j}~\Big\{G_{ij}^{\lessgtr}(E-\hbar\omega)\nonumber\\&+G_{ij}^{\lessgtr}(E+\hbar\omega)\Big\}
A_{\eta}(\mathbf{R}_{i},\hbar\omega) A_{\eta}(\mathbf{R}_{j},\hbar\omega),\\
&\mathbf{\Sigma}^{e\gamma-st,R}\approx \frac{1}{2}\Big(\mathbf{\Sigma}^{e\gamma-st,>}-\mathbf{\Sigma}^{e\gamma-st,<}\Big),
\label{eq:elphotse_stim}
\end{align}
with $\hbar\omega$ the photon energy, $\eta\in\{x,y,z\}$ the polarization component and the coupling matrix elements $\mathcal{M}^{e\gamma,\eta}_{i}\equiv -\big(\frac{e}{m_{0}}\big) \mathrm{p}_{cv}^{\eta}(\mathbf{R}_{i})$,
where $\mathrm{p}_{cv}$ is the inter-band momentum matrix element and
$\mathbf{R}_{i}$ the QD position. For electrons, the first term in curly brackets provides the generation, while the second term yields the stimulated emission. The attenuation of
the EM field relates to the absorption cross section (AC) $\mathcal{A}$ of the QD
array via the absorption coefficient $\alpha(\hbar\omega)=\rho_{QD}\mathcal{A}(\hbar\omega)$, where $\rho_{QD}$ is the QD density. For a single QD array or a very dilute QD system, the attenuation of the incident field can be neglected. The AC itself is computed based on the electronic structure from the net spectral absorption rate and the incident photon flux via $\mathcal{A}_{\eta}(\hbar\omega)\equiv \mathcal{R}_{abs,net}^{\eta}(\hbar\omega)/\Phi^{\gamma}_{0\eta}(\hbar\omega)$, with:
\begin{align}
\mathcal{R}_{abs,net}^{\eta}(\hbar\omega)=&\int \frac{dE}{2\pi\hbar}\sum_{i,j}\Big[G^{>}_{ij}(E)\Sigma^{e\gamma-st,<}_{ji}(E,\hbar\omega)\nonumber\\
&-G^{<}_{ij}(E)\Sigma^{e\gamma-st,>}_{ji}(E,\hbar\omega)\Big], \label{eq:absrate}
\end{align}
and $\Phi_{0\eta}^{\gamma}(\hbar\omega)=2 n_{r} c_{0}\varepsilon_{0}\hbar^{-1}\omega |A_{\eta}(\mathbf{R}_{0},\hbar\omega)|^2$, where a background refractive index $n_{r}$ is assumed, which for the numerical evaluations is set to $n_{r}=3.63$.
This yields the following expression for the AC in terms of the photon self-energy $\Pi$:
\begin{align}
\mathcal{A}_{\eta}(\hbar\omega)=\frac{c_{0}}{2 n_{r}\omega}\sum_{i,j}\Re\mathrm{e}\{i\hat{\Pi}_{ji,\eta\eta}(\hbar\omega)\}.\label{eq:abscrossect}
\end{align} 
Expression \eqref{eq:abscrossect} considers the net absorptance,
i.e., including stimulated emission, by the definition of 
$\hat{\Pi}\equiv\Pi^{>}-\Pi^{<}$ from the photon SE
components: 
\begin{align}
\Pi_{ij,\eta\eta'}^{\lessgtr}(E)=-i\hbar\mu_{0} \mathcal{M}^{e\gamma,\eta}_{i}\mathcal{M}^{e\gamma,\eta'}_{j}\mathcal{P}_{ij}^{\lessgtr}(E),
\end{align}
where the electron-hole polarization function elements are:
\begin{align}
\mathcal{P}_{nm}^{\lessgtr}(E)=\int\frac{dE'}{2\pi\hbar}G^{\lessgtr}_{nm}(E')G^{\gtrless}_{mn}(E'-E),
\end{align} 
and which is hence related to the spectral function
$\hbar\hat{\mathcal{P}}/2\pi$ of occupied electron-hole pairs ($\hat{\mathcal{P}}\equiv\mathcal{P}^{>}-\mathcal{P}^{<}$)\cite{ae:prb_11}. At global quasi-equilibrium conditions, where the carrier populations are fixed by global quasi-Fermi levels (QFL) $\mu_{b}$ ($b=c,v$), the fluctuation-dissipation theorem provides the relations: 
\begin{align}
\mathbf{G}_{b}^{\lessgtr}(E)=i\hat{\mathbf{G}}_{b}(E)[f(E-\mu_{b})-1/2\pm 1/2]
\end{align}
where: 
\begin{align}
\hat{\mathbf{G}}\equiv& i\left(\mathbf{G}^{>}-\mathbf{G}^{<}\right)=i\left(\mathbf{G}^{R}-\mathbf{G}^{R,\dagger}\right)=-2\Im \mathrm{m} \mathbf{G}^{R}\label{eq:carrspecfun}
\end{align} 
are the spectral functions of the charge carriers. Far away from degeneracy and under low intensity excitation, band filling effects can be neglected, and the electron-hole spectral function is proportional to the Joint Density of States (JDOS) of the system, which in turn can be written in terms of the spectral
functions of the individual charge carriers as follows:
\begin{align}
\mathcal{J}_{cv}(E)=&\int \frac{dE'}{4\pi^2} ~\mathrm{tr}\left\{\hat{\mathbf{G}}_{c}(E')\hat{\mathbf{G}}_{v}(E'-E)\right\}\\
&\equiv \sum_{i}\mathcal{J}_{cv,i}(E),
\end{align}
defining the local joint density of states (LJDOS) as:
\begin{align}
\mathcal{J}_{cv,i}(E)=&\int \frac{dE'}{4\pi^2}\sum_{k}\hat{G}_{c,ik}(E')\hat{G}_{v,ki}(E'-E)\\
\equiv& \sum_{k}\mathcal{J}_{cv,ik}(E).\label{eq:nonloc}
\end{align}
While the LJDOS is directly related to the local absorption coefficient, the last expression defines the contributions to the (L)JDOS according to non-locality, i.e., according to the extent of spatially off-diagonal elements of the carrier Green's functions considered in \eqref{eq:nonloc}. Fig.~\ref{fig:nonloc} shows the evolution of the JDOS with increasing degree of non-locality: while - remarkably - the sum rule is observed in all cases, the JDOS deviates strongly from the full result for restricted non-locality, as observed previously for other nano-systems\cite{pourfath:09,ae:jpe_14}.  
As for the DOS, the JDOS converges with increasing number of QDs to that of the infinite tight-binding chain for two coupled single orbital bands. This is displayed in the inset of Figure \ref{fig:nonloc}, where the JDOS provided by the NEGF formalism for a system of 300 coupled and selectively contacted QD (open circles) is shown together with the result for the infinite 1D TB chain (solid line), which in complete analogy to \eqref{eq:1dtb_dos} reads
\begin{align}
\mathcal{J}_{cv,\infty}(E)=\left[2\pi t_{cv}\sqrt{1-\left(\frac{E+i\eta_{cv}-\varepsilon_{cv}}{2t_{cv}}\right)^2}\right]^{-1}
\end{align}
where $t_{cv}=|t_{c}|+|t_{v}|$, $\varepsilon_{cv}=\varepsilon_{c}-\varepsilon_{v}$ and $\eta_{cv}=\eta_{c}+\eta_{v}$.

\begin{figure}[t]
\begin{center}
\includegraphics[width=0.4\textwidth]{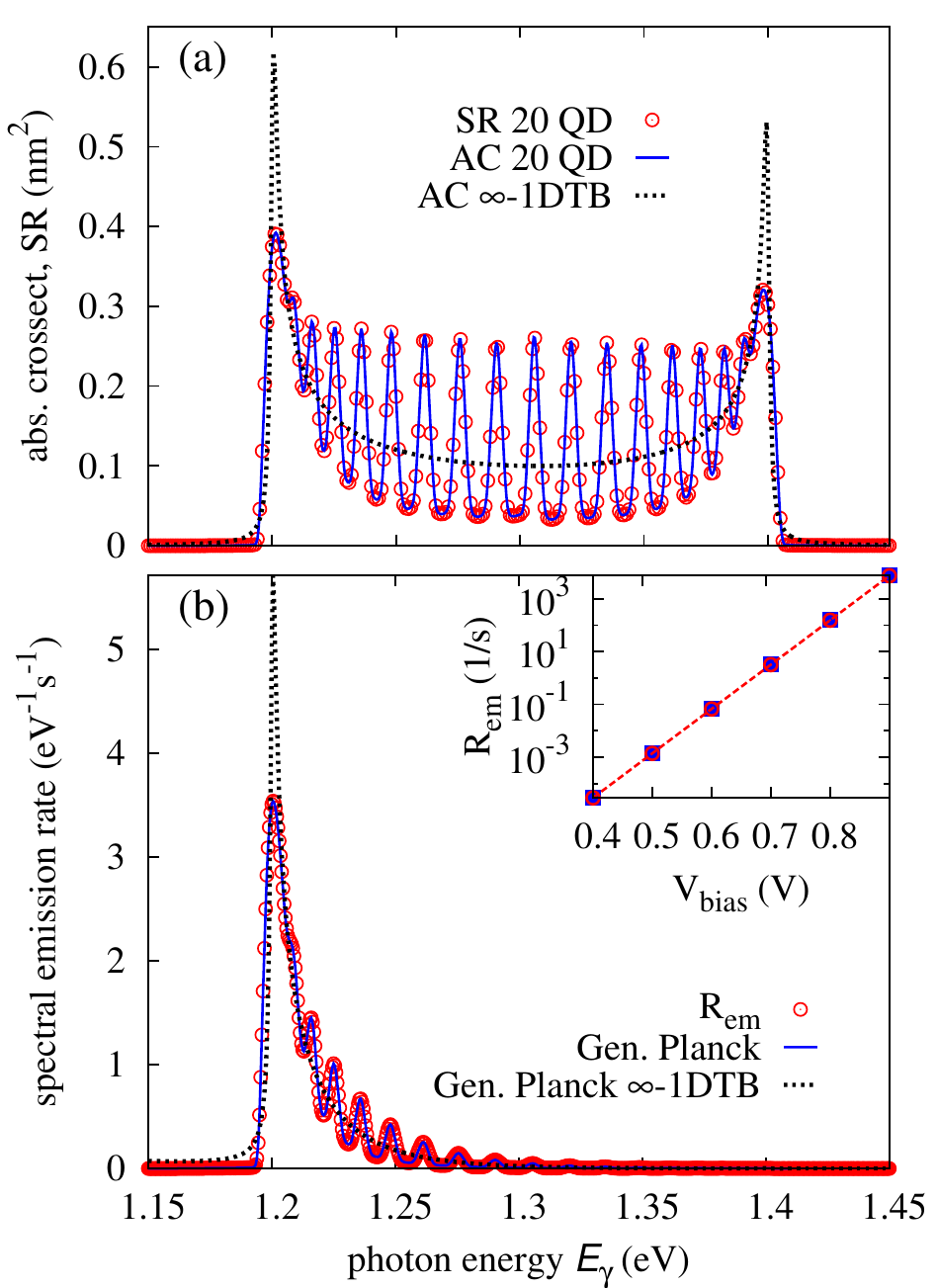}
\caption{(color online)(a) Absorption cross section (AC) and Spectral Response (SR) of the 20 QD cell as obtained from the NEGF, in comparison with the AC of the infinite 1D TB model; (b) Spectral emission rate for the 20 QD cell at a terminal bias voltage $V_{\textrm{bias}}\equiv\mu_{R}-\mu_{L}=0.6$ V as derived from the NEGF formalism (empty dots) as well as from the Generalized Planck law using the NEGF absorption (solid line) or the infinite 1D TB model (dotted line). The inset shows the integrated emission rate as a function of $V_{\textrm{bias}}$, with open symbols for the values given by the current expression \eqref{eq:current_NEGF} and filled squares for the values obtained from the global emission rate \eqref{eq:emrate}.  
\label{fig:JDOS_emspect}}
\end{center}
\end{figure}

The spectral response (SR) $\mathcal{S}$ of the system is defined in terms of the total current $I^{tot}$ for a given monochromatic illumination at photon energy $E_{\gamma}=\hbar\omega$ in the form 
\begin{align}
\mathcal{S}(\hbar\omega)\equiv I^{tot}(\hbar\omega)/\{e\Phi^{\gamma}_{0}(\hbar\omega)\}. 
\end{align}
The total charge current $I^{tot}=I^{e}+I^{h}$ is directly obtained from the charge carrier currents as computed within the NEGF formalism, e.g., via:
\begin{align}
I^{e}_{i}=\frac{e}{\hbar}\int
\frac{dE}{\pi}~2\Re\mathrm{e}\left\{H_{0,ii+1}G^{<}_{i+1i}(E)\right\}\label{eq:current_NEGF}
\end{align}
for the electron current in the conduction band between QDs at positions $\mathbf{R}_{i}$ and
$\mathbf{R}_{i+1}$\cite{lake:06}. A similar expression exists
for the hole current $I^{h}$ in the valence band in terms of the
hole NEGF component $\mathbf{G}^{>}$. The total current can be
compared to the photocurrent as given by the absorptance and the
photon flux via the spectral relation for the generation current:
\begin{align}
I_{\mathcal{A}}(\hbar\omega)=e\sum_{\eta}\Phi_{0\eta}^{\gamma}(\hbar\omega)\mathcal{A}_{\eta}(\hbar\omega)\equiv e 
\mathcal{R}_{abs,net}(\hbar\omega).
\end{align}
The extraction efficiency for the photogenerated charge carriers is
then given by the ratio $\eta_{ext}=I^{tot}/I_{\mathcal{A}}$. In the situation displayed in Fig.~\ref{fig:JDOS_emspect}(a), the SR equals the absorption cross section $\mathcal{A}$, which corresponds to unit extraction efficiency $\eta_{ext}=1$, i.e., $I^{tot}=I_{\mathcal{A}}$.

\subsection{Radiative recombination and dark current}

At the radiative limit, the dark current-voltage characteristics of
the selectively contacted QD array solar cell is determined by the
current due to radiative recombination of charge carriers injected
at the contacts. For the description of the associated spontaneous
emission process, consideration of the photon Green's function of
the entire system is required to include all the optical modes with
finite coupling. Under the assumption of an optically homogeneous
medium, use of the free field photon propagator results in the SE\cite{ae:jpe_14}
\begin{align}
\Sigma_{ij}^{e\gamma-sp,\lessgtr}(E)\approx&\frac{\mu_{0}n_{r}}{\pi c_{0}}\bar{\mathcal{M}}^{e\gamma}_{i}\bar{\mathcal{M}}^{e\gamma}_{j}\int_{0}^{\infty}\frac{dE'}{2\pi\hbar}E' G^{\lessgtr}_{ij}(E\pm E'),\label{eq:spontem_se}
\end{align}
where $\bar{\mathcal{M}}$ are the polarization-averaged coupling elements. In
terms of the photon SE,
the polarization-averaged spectral emission rate of the QD array
acquires the form
\begin{align}
\bar{\mathcal{R}}_{em}(\hbar\omega)=\frac{n_{r}\omega}{2 \pi^2\hbar c_{0}}\sum_{i,j}\Re\textrm{e}\{i\bar{\Pi}_{ji}^{<}(\hbar\omega)\},\label{eq:emrate}
\end{align}
which is obtained by using $\Sigma^{e\gamma-sp}$ in
the expression for the emission rate, in analogy to
Eq.~\eqref{eq:absrate}. For valid assumption of a \emph{global} quasi-equilibrium, the \emph{Kubo-Martin-Schwinger} relation between the
components of the polarization function\cite{pereira:98,richter:08},
$\mathcal{P}^{<}(\hbar\omega)=\mathcal{P}^{>}(\hbar\omega)e^{-\beta(\hbar\omega-\mu_{cv})}$,
with $\beta=(k_{B}T)^{-1}$ and $\mu_{cv}=\mu_{c}-\mu_{v}$ the
quasi-Fermi level splitting (QFLS),
can be used in Expr.~\eqref{eq:abscrossect} and
\eqref{eq:emrate} to derive the
\emph{Generalized Planck} law\cite{wuerfel:82} via the connection:\footnote{Strictly, this relation between absorption at zero bias an emission under finite QFLS is only valid if neither JDOS nor the optical matrix elements $\mathcal{M}^{e\gamma}$ are modified at finite bias voltage, e.g., due to a change in orbital overlap or local band bending.}
\begin{align}
\bar{\mathcal{R}}_{em}(\hbar\omega)=\bar{\mathcal{A}}(\hbar\omega)\frac{(\hbar\omega)^2 n_{r}^2}{\pi^2\hbar^3 c_{0}^2}\Big\{e^{\beta(\hbar\omega-\mu_{cv})}-1\Big\}^{-1}.
\end{align}

To obtain the QFLS, the individual QFL $\mu_{c}$ and $\mu_{v}$ of
electrons and holes in conduction and valence bands need to be
determined. They can be defined via the carrier density, and the
spectral function \eqref{eq:carrspecfun}, under the assumption of quasi-equilibrium
conditions. An energy
independent value of the QFL is extracted from the
density determined via energy integration of $G^{\lessgtr}$. Figure \ref{fig:JDOS_emspect}(b) shows the comparison of the electroluminescence spectra as obtained from the NEGF via the above formalism for the 20 QD system at a terminal bias voltage $V_{\textrm{bias}}=\mu_{R}-\mu_{L}=0.6$ V (empty dots) with the Generalized Planck emission for AC determined either via NEGF for the same finite system (solid line) or for the infinite TB chain (dotted line). The effective emission edge of the 20 QDs cell is close to that of the infinite 1D TB model, as expected from the convergence of the spectral width of the JDOS seen in Fig.~\ref{fig:JDOS_emspect}(a). The linear increase of the QFLS  with $V_{\textrm{bias}}$ results in an exponential growth of the emission rate, as displayed in the inset of Fig.~\ref{fig:JDOS_emspect}(b).

\subsection{Carrier extraction efficiency\label{subsec:exteff}}

\begin{figure}[t!]
\begin{center}
\includegraphics[width=0.45\textwidth]{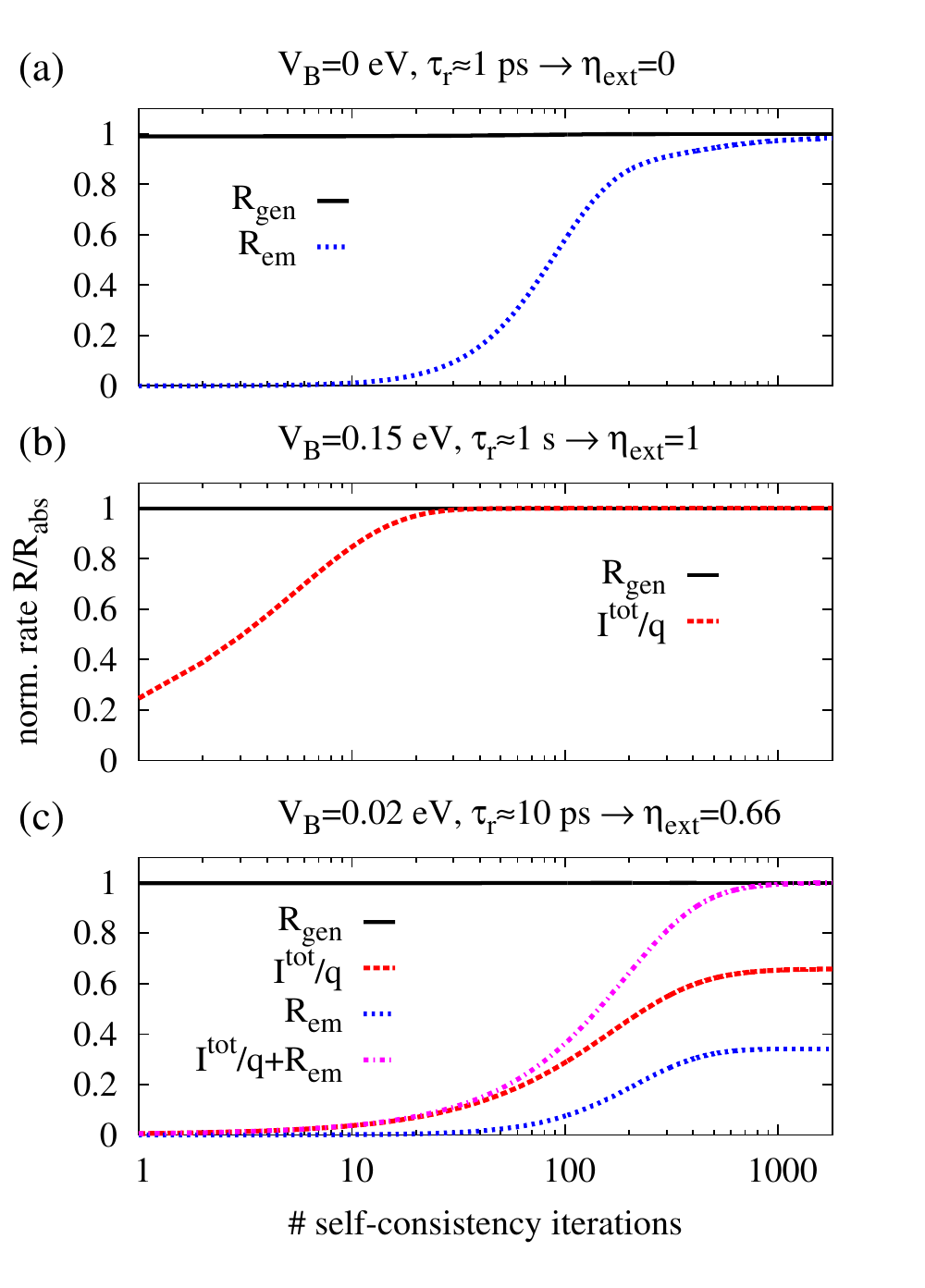}
\caption{(color online) Evolution during the NEGF self-consistency iteration of the energy-integrated rates for charge carrier generation ($R_{gen}$), radiative recombination ($R_{em}$) and extraction ($I^{tot}/q$), normalized to the absorption rate given by the incident photon flux and the absorptance, for different regimes of operation: (a) Optical extraction only ($\rightarrow \eta_{ext}=0$), i.e., photoluminescence at closed contacts  ($V_{c}=V_{v}\equiv V_{B}=0$) and a recombination lifetime $\tau_{r}\approx 1$ ps; (b) Unit carrier extraction at electrodes ($\rightarrow \eta_{ext}=1$) due to strong dot-contact coupling $V_{cR}=V_{vL}\equiv V_{B}=0.15$ eV and long recombination lifetime $\tau_{r}\approx 1$ s; (c) Incomplete carrier extraction ($\rightarrow 0<\eta_{ext}<1$) for recombination lifetimes ($\tau_{r}\approx 10$ ps) comparable to the escape time at finite, but low contact coupling  $V_{cR}=V_{vL}\equiv V_{B}=0.02$ eV. In all cases, an inter-dot coupling of $|t_{c/v}|=0.03/0.02$ eV is used.
\label{fig:scit_conv}}
\end{center}
\end{figure}

Reduced carrier extraction efficiencies ($\eta_{ext}<1$) occur when the charge recombination rate reaches a level that is comparable to the extraction rate. In the NEGF formalism, these two competing rates are obtained from expressions similar to Eq.~\eqref{eq:absrate}, using the out-scattering component of either; the self-energy of the recombination process (i.e., of the spontaneous emission \eqref{eq:spontem_se} in the case of the radiative limit), or the contact self-energy \eqref{eq:contactse}, since the total current equals the sum of electron and hole terminal currents,
\begin{align}
I_{B}=q\int \frac{dE}{2\pi\hbar}\mathrm{tr}\Big[\boldsymbol{\Sigma}^{>B}(E)\boldsymbol{G}^{<}(E)-\boldsymbol{\Sigma}^{<B}(E)\boldsymbol{G}^{>}(E)\Big],
\end{align}
where the integration is over both conduction and valence bands. Since the total current is constant over the whole device, $I_{B}=I^{tot}$ holds. The radiative and extraction lifetimes $\tau_{rad,ext}$ are directly proportional to the imaginary part of the corresponding self-energy, i.e., to the broadening function 
\begin{align}
\Gamma_{x}\equiv i\left(\Sigma^{>}_{x}-\Sigma^{<}_{x}\right)=i\left(\Sigma^{R}_{x}-\Sigma^{R\dagger}_{x}\right),
\end{align}
via $\tau_{x}=\hbar/\Gamma_{x}$ ($x=rad,~ext$). Hence, sub-unit carrier extraction efficiencies can result from either low rates of escape into contacts or increased rates of recombination within the absorber. The relation between carrier recombination and extraction is reflected in the evolution during the self-consistency iteration of GF and SE of the recombination rate or terminal current towards the steady-state generation rate or current, respectively. This is shown in Fig.~\ref{fig:scit_conv} for the two extreme cases of (a) photoluminescence $\eta_{ext}=0)$, i.e., vanishing carrier extraction due to closed contacts ($V_{c}=V_{v}\equiv V_{B}=0$), assuming a recombination lifetime $\tau_{r}\approx 1$ ps\footnote{In order to reproduce lifetimes of recombination processes that compete with carrier extraction, the radiative recombination is enhanced by a corresponding factor in the self-energy of the spontaneous emission}, and (b) perfect photocarrier extraction ($\eta_{ext}=1)$ for large contact coupling ($V_{cR}=V_{vL}\equiv V_{B}=0.15$ eV) and long carrier lifetime $\tau_{r}=1$ s. In the intermediate case (c), smaller values of contact coupling ($V_{cR}=V_{vL}\equiv V_{B}=0.02$ eV) and radiative lifetime ($\tau_{r}\approx 10$ ps) lead to an incomplete carrier extraction ($0<\eta_{ext}<1$) due to recombination losses. In all cases, an inter-dot coupling of $|t_{c/v}|=0.03/0.02$ eV is used, and the rates are normalized to the absorption rate $R_{abs}=\Phi_{0}\cdot\mathcal{A}$.

\section{Numerical results and discussion\label{sec:results}}

\begin{figure}[t] 
\begin{center}
\includegraphics[width=0.5\textwidth]{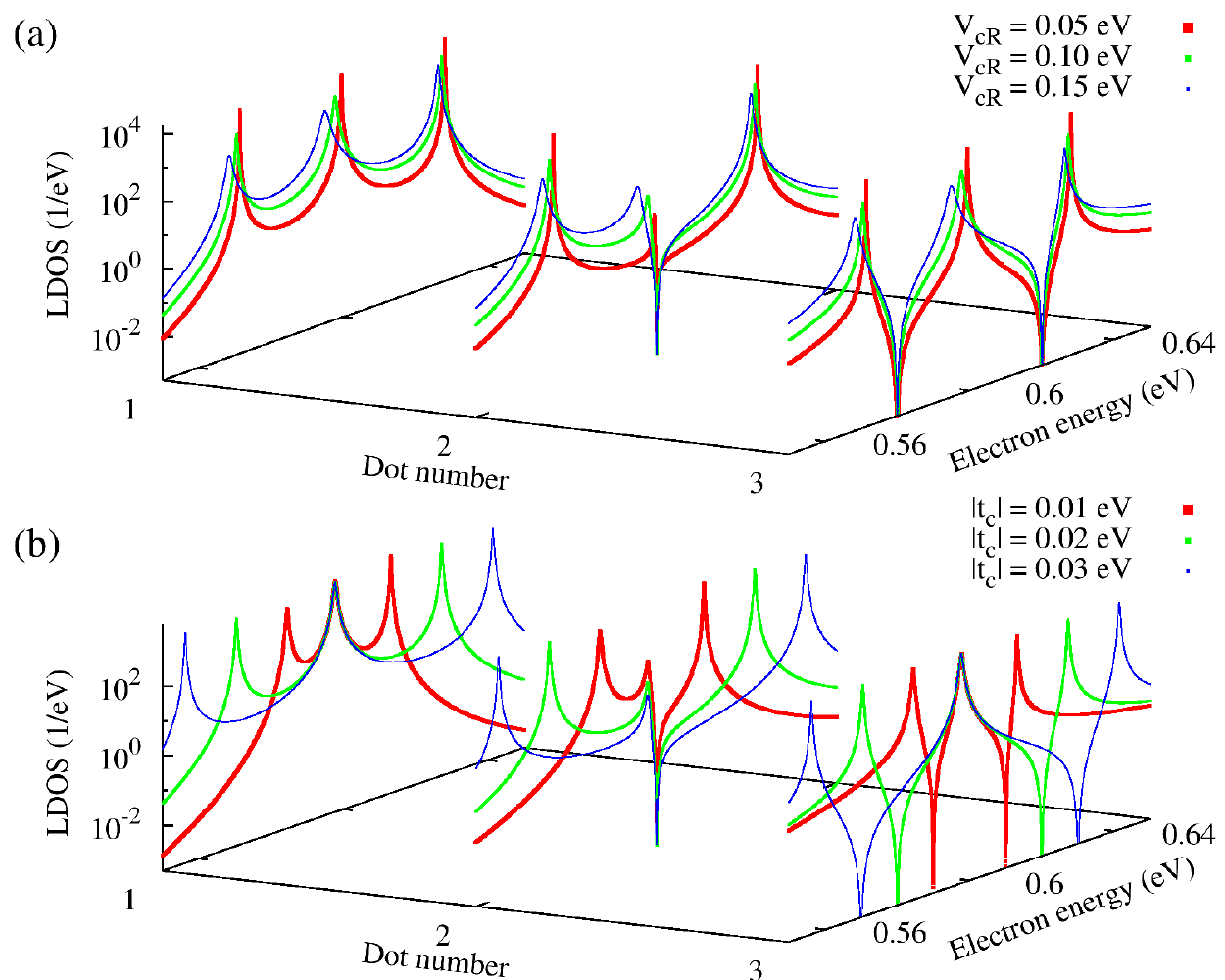}
\caption{(color online) Local density of electron states of a right-contacted 3 QD 
system for varying values (a) of the contact coupling $V_{cR}$ at fixed inter-dot coupling $|t_{c}|=0.02$ eV and (b) of the inter-dot coupling $|t_{c}|$ at fixed contact coupling $V_{cR}=0.1$ eV. The LDOS far from the contact acquires a regular form with
peak multiplicity determined by the number of QD. The spectral shape
of the LDOS in the direct vicinity of the contact exhibits the
asymmetric Fano-type signatures of the coupling of a discrete level
system to the continuum of states in the bulk electrode. The effects
of the contact coupling are a level broadening and a red-shift that both increase with the
coupling strength $V_{cR}$. The energy separation of the resonances, on
the other hand, is directly proportional to the inter-dot coupling
$|t_{c}|$. \label{fig:LDOS_3qd}}
\end{center} 
\end{figure}

\begin{figure*}[t!]
\begin{center}
\includegraphics[width=1.0\textwidth]{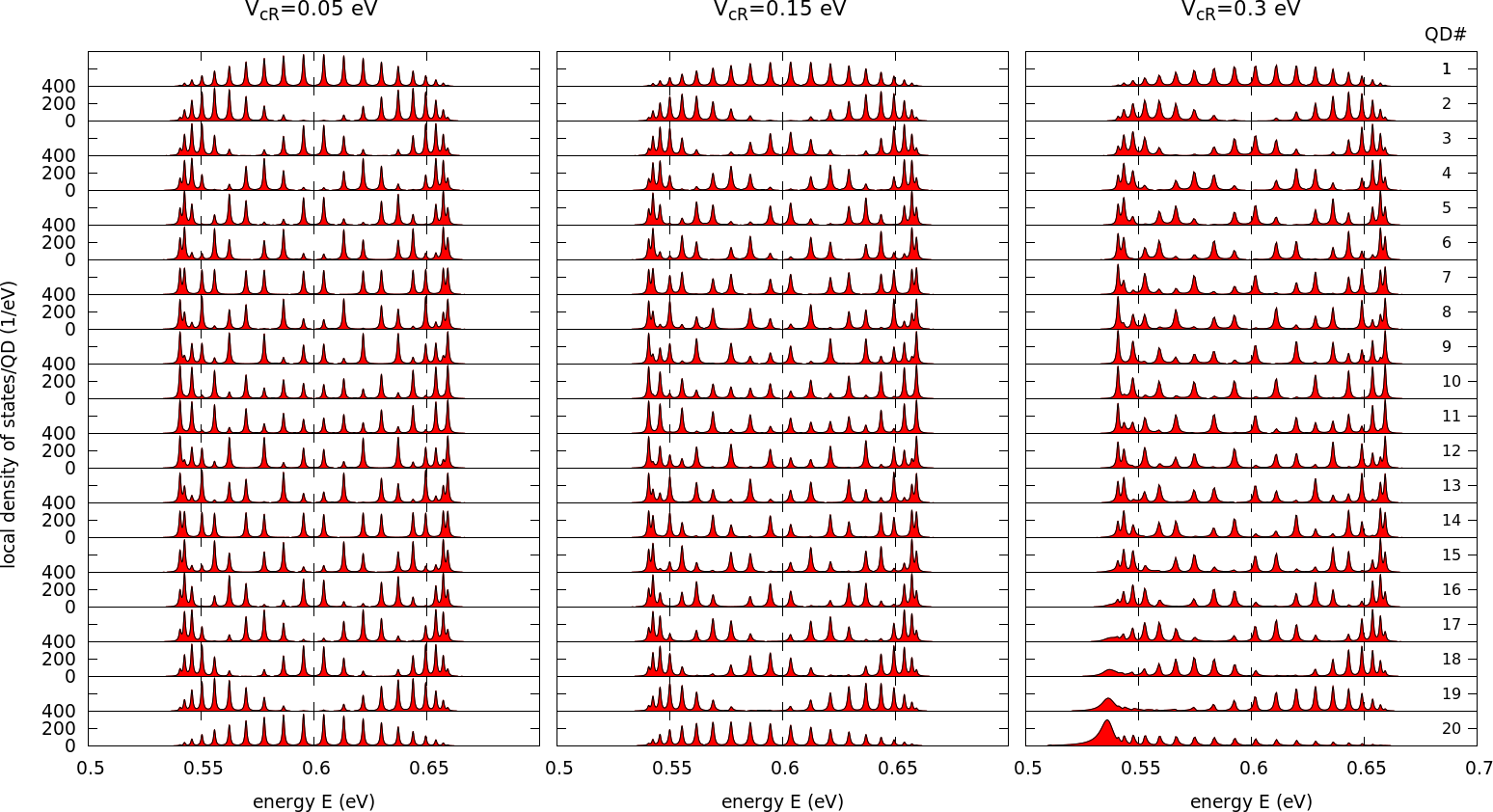}\\
\caption{(color online) LDOS of a selectively contacted 20 QD array resolved on the individual dot sites and displayed for different values of the contact coupling $V_{cR}=0.05/0.15/0.3$ eV, at a fixed inter-dot coupling of $|t_{c}|=0.03$ eV. While for the weakest coupling, the DOS is almost symmetric both along the array and with respect to the central level energy at $\varepsilon_{c}=0.6$ eV, the asymmetry induced to the  real part of the contact self energy is already visible at intermediate coupling and results in the formation of a bound interface state  below the low energy edge of the miniband at the contact layer. In addition, there is the broadening of the DOS increasing with coupling strength, which originates in the imaginary part of the contact self energy.  
 \label{fig:LDOS_20qd}}
\end{center} 
\end{figure*}

On the basis of this general analysis, the specific impact of the dot-contact and inter-dot coupling on the density of states, spectral response and radiative recombination characteristics can now be discussed. This is first done for the situation in which photocarrier extraction is much faster than recombination ($\eta_{ext}\rightarrow 1$), and then for situations where carrier extraction competes with radiative recombination ($\eta_{ext}<1$). All the results under illumination are obtained assuming a monochromatic spectrum with $E_{\gamma}=1.3$ eV at an intensity of $I_{\gamma}=10$ kW/m$^2$ and normal incidence ($A_{\eta}=A_{0}\delta_{\eta,x}$). The corresponding momentum matrix element is approximated via $p_{cv}^{x}\approx \bar{p}_{cv}=\sqrt{p_{cv}^2/3}=\sqrt{E_{P} m_{0}/6}$, with the Kane energy $E_{P}=2.6$ eV.\cite{bartnik:07} 

\subsection{$\eta_{ext}\rightarrow 1$}
Figure \ref{fig:LDOS_3qd}(a) displays the LDOS of electrons in a 3 QD
chain contacted from the right, at a fixed inter-dot coupling $|t_{c}|=0.02$ eV and varying dot-contact coupling strength. The hybridization of the discrete QDs
levels with the continuum of bulk electrode states introduces
Fano-type asymmetric features close to the contact, which gradually disappear
with increasing distance from the electrode. The main effect of the
presence of this contact is the broadening of the LDOS, especially
in the vicinity of the electrode, where the band width can be
strongly increased as shown in Fig.~\ref{fig:ldos_2b_eh}.
Since the electrode self-energy exhibits a real part, the magnitude of
$V_{cR}$ not only affects this broadening, but also induces a shift
in the resonance peaks. As can be inferred from
Fig.~\ref{fig:LDOS_3qd}(b), where the contact coupling is fixed at a value of $V_{cR}=0.1$ eV, the inter-dot coupling acts on the LDOS
mainly via the direct relation to the miniband width. These effects of the coupling parameters are retained in larger arrays. Figure \ref{fig:LDOS_20qd} shows the electronic LDOS of a 20 QD array resolved on the individual dot sites, for a fixed inter-dot coupling $|t_{c}|=0.03$ eV and different values of the dot-contact coupling. At weak coupling of $V_{B}=0.05$ eV, the LDOS is symmetric with respect to both the contacts and the central level energy, i.e., it is not affected by the coupling to the electrode states. For  an intermediate coupling of $V_{B}=0.15$ eV, there are visible contact induced effects such as broadening and a shifting of the DOS weight towards the lower energy band edge on QD sites in the vicinity of the contact. Finally, for strong coupling of $V_{B}=0.3$ eV, localized surface states appear below the lower band edge at the open contact, while the LDOS remains almost unaffected in the vicinity of the closed contact.

\begin{figure}[t]
\begin{center}
\includegraphics[width=0.46\textwidth]{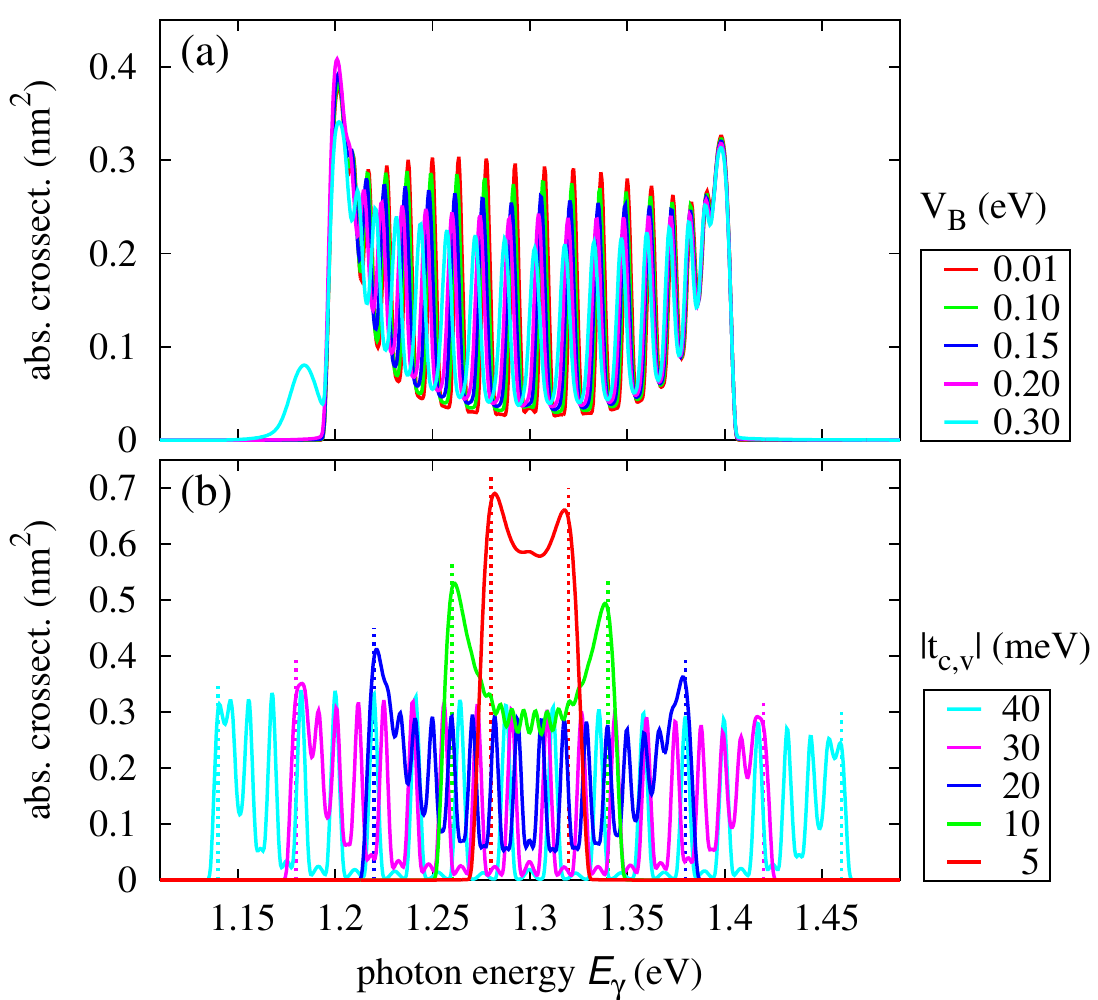}
\caption{(color online) The absorption cross section for the 20 QD cell is presented in (a) for various values of dot-contact couplings $V_{cR}=V_{vL}\equiv V_{B}$ at $|t_{c/v}|=0.03/0.02$ eV and in (b) for different inter-dot coupling parameters $t=|t_{c}|=|t_{v}|$ at $V_{cR}=V_{vL}\equiv V_{B}=0.05$ eV (dashed lines indicate band width of 1D TB chain).
\label{fig:JDOS_2}}
\end{center}
\end{figure}

\begin{figure}[t]
\begin{center}
\includegraphics[width=0.46\textwidth]{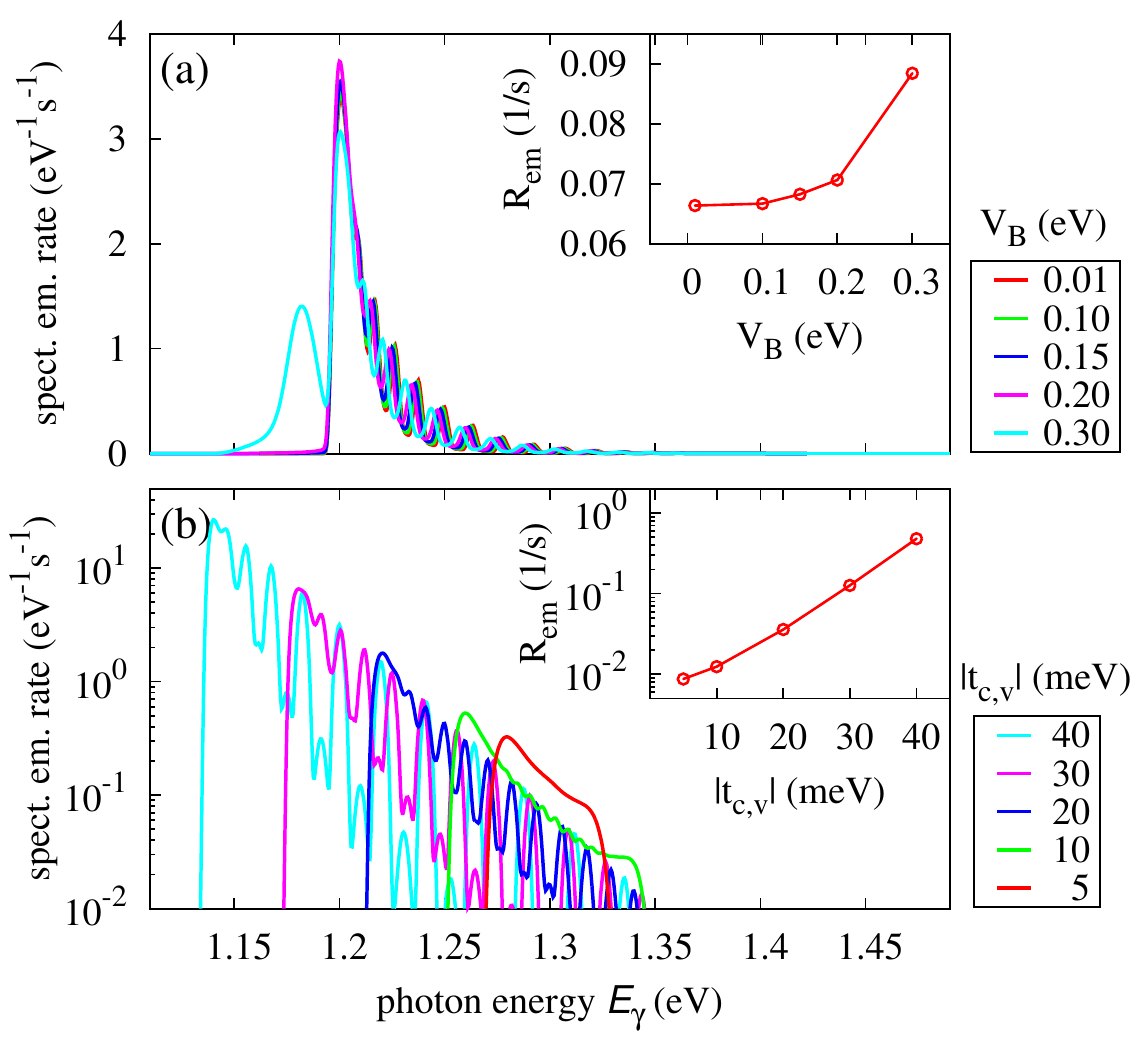}
\caption{(color online) Spectral emission rate for the 20 QDs at a terminal bias voltage $V_{\textrm{bias}}=0.6$ V, in (a) for different dot-contact couplings and in (b) for different inter-dot couplings. The inset displays the corresponding variation of the energy-integrated emission rate.
\label{fig:emspect_2}}
\end{center}
\end{figure}

The dependence of absorption and emission characteristics on the dot-contact and the inter-dot coupling are displayed in Figs.~\ref{fig:JDOS_2} and  \ref{fig:emspect_2}, respectively. In Fig.~\ref{fig:JDOS_2}(a), the spread in energy of the individual dot contributions to the absorption cross section is determined by the contact induced broadening. The width of the AC, on the other hand, is determined by the inter-dot coupling in analogy to the DOS of the individual bands, as shown in Fig.~\ref{fig:JDOS_2}(b). It converges with increasing number of dots to that of the single orbital TB chain (dashed lines).
Figures \ref{fig:emspect_2} (a) and (b) show the impact of contact and inter-dot coupling strength on
the levels of radiative dark current generation. Up to moderate coupling strength ($V_{B}\leq 0.2$ eV), the effective
gap is only marginally affected  by the contact coupling, and there is just
a slight increase of the dark current with stronger contact
hybridization due to the contact induced smearing and renormalization (i.e., red-shifting) of the emission edge, as
displayed in Fig.~\ref{fig:emspect_2}(a). At larger $V_{B}$, the surface states modify the optical spectra, most pronouncedly that of the emission, with considerable increase in the integrated emission rate. On the other hand,
Fig.~\ref{fig:emspect_2}(b) reveals the exponential increase of
the recombination with inter-dot coupling strength due to the linear
decrease of effective band gap
$\varepsilon_{cv}^{eff}=\varepsilon_{cv}^{0}-2(|t_{c}|+|t_{v}|)$, i.e.,
the decrease of the energy of the dominant contribution to the overall emission
process.

\subsection{$\eta_{ext}<1$}

In the following, the carrier lifetime is limited to a level comparable to that of the extraction by the presence of effective charge carrier recombination processes, which can result in sub-unit extraction efficiency, corresponding to recombination prior to extraction. The NEGF formalism used for the evaluation of extraction efficiency is based on a steady state picture of carrier transport, where the different lifetimes do not appear explicitly. However, as outlined in Sec.~\ref{subsec:exteff} and illustrated in Fig.~\ref{fig:scit_conv}, the time evolution of the interacting system is mimicked by the self-consistency iteration of carrier Green's functions and interaction self-energies towards the stationary state. 

\begin{figure}[t!]
\begin{center}
\includegraphics[width=0.48\textwidth]{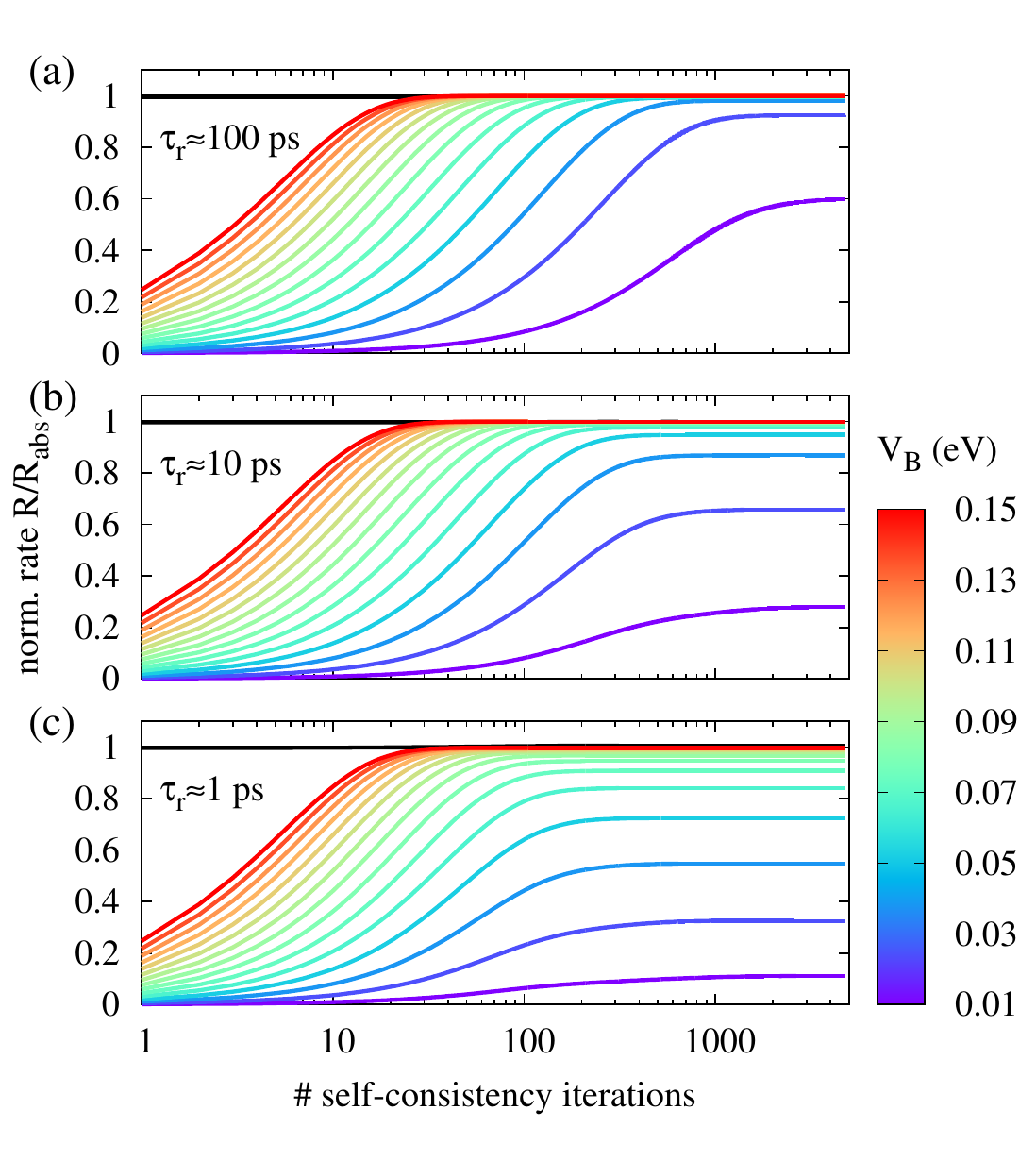}
\caption{(color online) 
\label{fig:scit_conv_varVc} Convergence of the terminal current in the NEGF self-consistency iteration for different values of contact coupling $V_{B}$, fixed inter-dot coupling $|t_{c/v}|=0.03/0.02$ eV and recombination lifetimes of (a) $\tau_{r}\approx 100$ ps, (b) $\tau_{r}\approx 10$ ps and (c) $\tau_{r}\approx 1$ ps.}   
\end{center}
\end{figure}

\begin{figure}[t!]
\begin{center}
\includegraphics[width=0.45\textwidth]{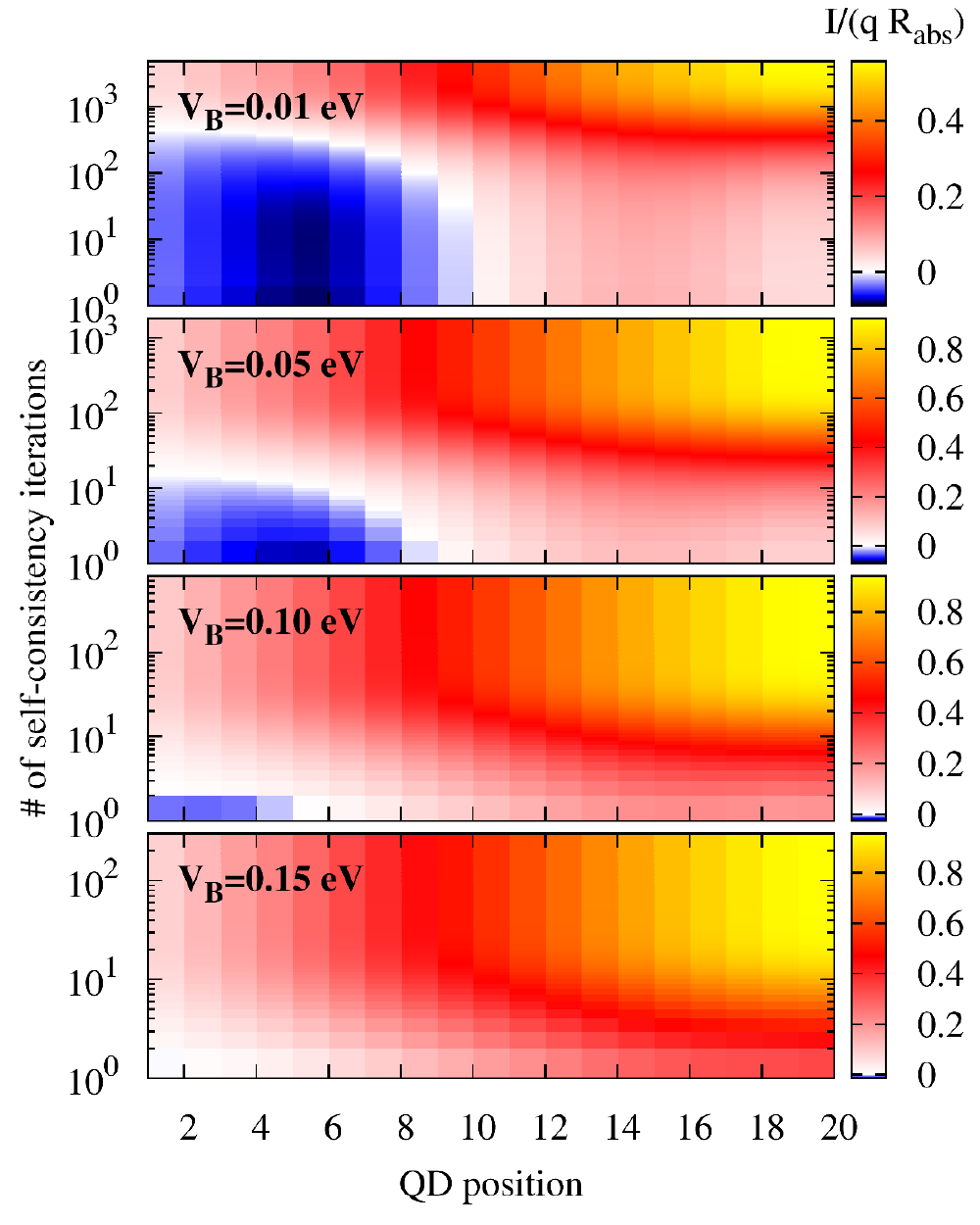}
\caption{(color online) 
\label{fig:scit_currevol}Convergence of the local current resolved on individual dot position and for different values of the contact coupling $V_{B}$, revealing a pronounced dependence of photocurrent rectification on the size of the coupling, which is related to the extraction rate as well as to reflection of photogenerated charge carriers.}  
\end{center}
\end{figure}

In Fig.~\ref{fig:scit_conv_varVc}, the convergence of the terminal current normalized to the absorption rate is shown for a 20 QD system with $|t_{c/v}|=0.03/0.02$ eV, $V_{B}\in [0.01,0.15]$ eV\footnote{The upper bound to $V_{B}$ is chosen in order to exclude spurious effects of surface states.} and an effective recombination lifetime of (a) $\tau_{r}\approx 100$ ps, (b) $\tau_{r}\approx 10$ ps and (c) $\tau_{r}\approx 1$ ps. For large dot contact coupling, unit extraction efficiency is reached after a small number of iterations and independent of lifetime. The smaller the lifetime, the larger the critical contact coupling at which the system converges to a stationary state with sub-unit extraction efficiency, and the sooner this convergence is achieved. For a qualitative explanation of this behavior, the convergence of the current is displayed in Fig.~\ref{fig:scit_currevol} in spatial resolution. This reveals a photocurrent rectification process whose speed depends on the size of the dot-contact coupling: the larger the coupling, the larger the electron "sink" and the less reflection at the open contact, i.e., the smaller the negative current contributions corresponding to the fraction of charge carriers traveling towards the ''wrong'' contact.

\begin{figure}[t]
\begin{center}
\includegraphics[width=0.48\textwidth]{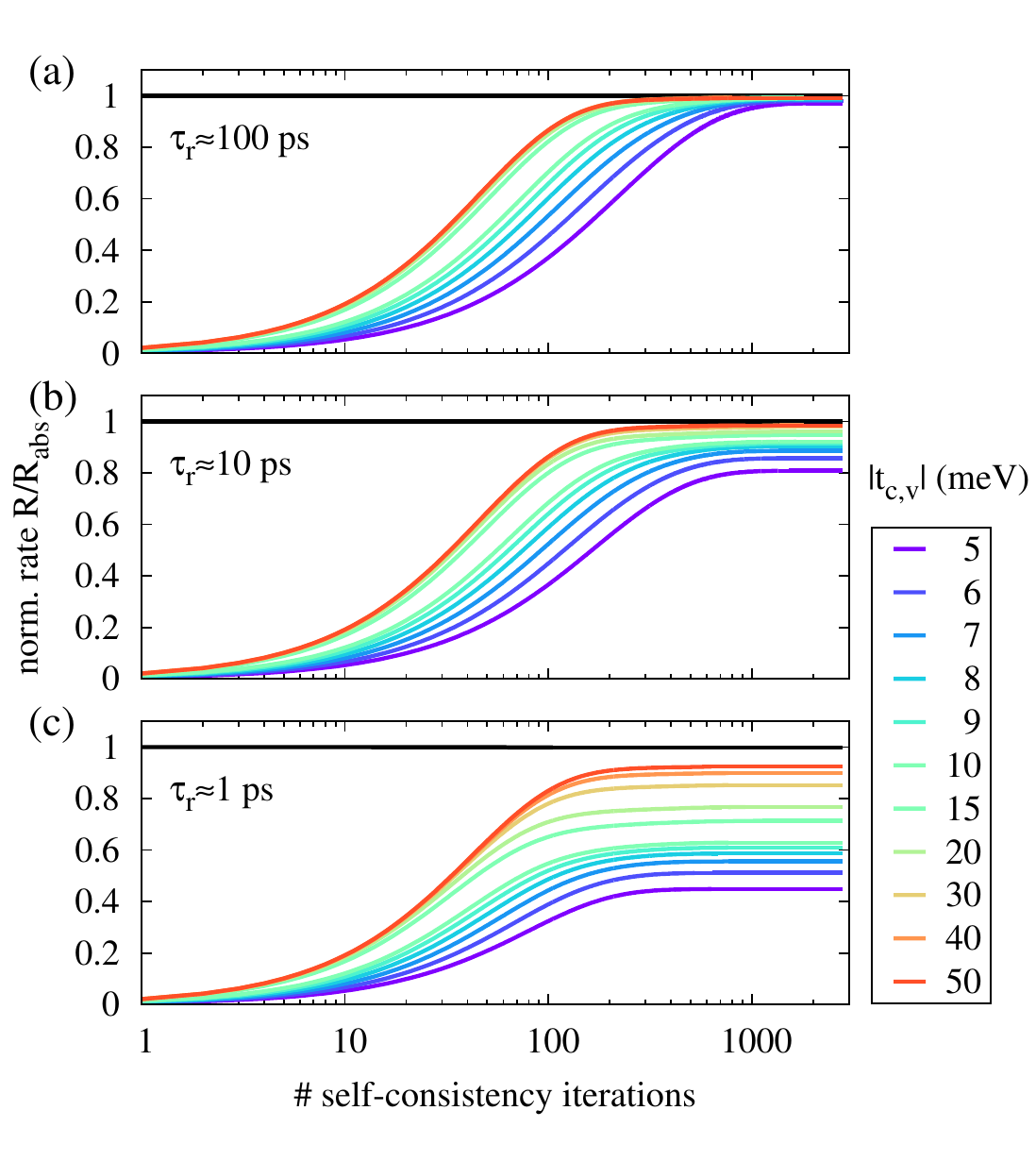}
\caption{(color online) 
\label{fig:scit_conv_vartc}Convergence of the terminal current in the NEGF self-consistency iteration for different values of inter-dot coupling $|t_{c,v}|$, fixed contact coupling $V_{B}=0.05$ eV and recombination lifetimes of (a) $\tau_{r}\approx 100$ ps, (b) $\tau_{r}\approx 10$ ps and (c) $\tau_{r}\approx 1$ ps.}
\end{center}
\end{figure}

The convergence behavior of the terminal current with respect to varying inter-dot coupling $|t_{c,v}|\in [5,50]$ meV for fixed contact coupling $V_{B}=0.05$ eV and the same values for recombination lifetimes as in Fig.~\ref{fig:scit_conv_varVc} is displayed in Fig.~\ref{fig:scit_conv_vartc}. As compared to the effect of varying $V_{B}$, the spread induced by varying $t_{c,v}$ is less pronounced, and at a recombination lifetime of $\tau_{r}\approx 100$ ps, unit extraction efficiency is achieved for all values of the inter-dot coupling considered.

\begin{figure}[t!]
\begin{center}
\includegraphics[width=0.47\textwidth]{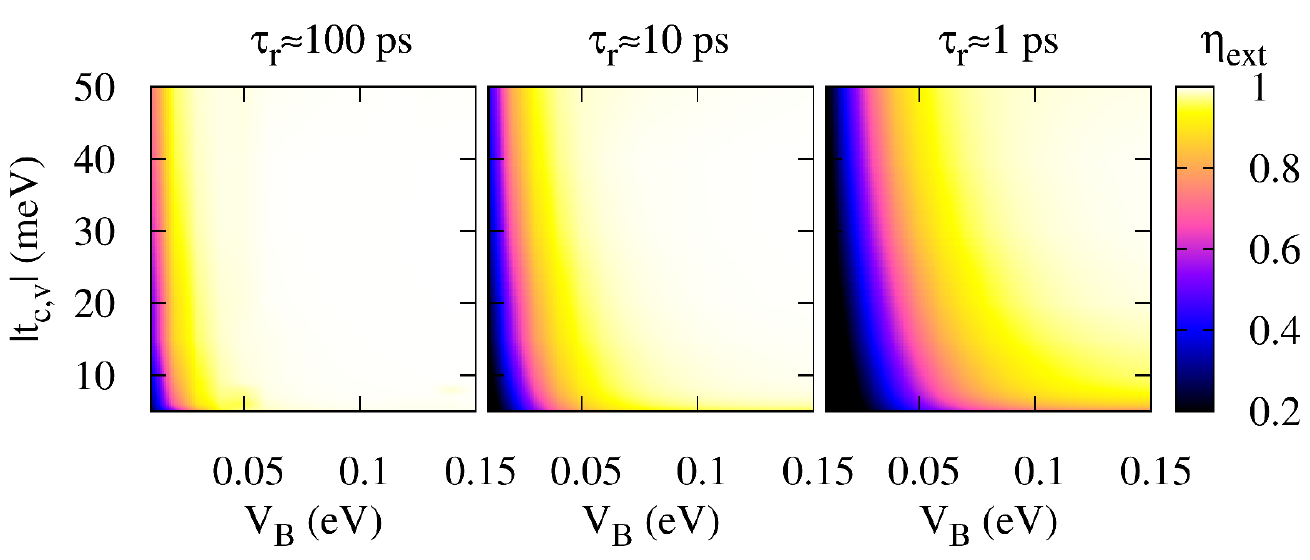}
\caption{(color online) Carrier extraction efficiency for configurations of inter-dot coupling $|t_{c,v}|\in [5,50]$ meV and dot-contact coupling $V_{B}\in [0.01,0.15]$ eV and different recombination lifetimes. At all lifetimes, the predominant effect on the extraction efficiency originates in the dot-contact coupling.
\label{fig:extracteff_map}}
\end{center}
\end{figure}

Finally, the photocarrier extraction efficiency is evaluated for the whole configurational parameter space $\{V_{B},|t_{c,v}|\}\in [10,150]\otimes[5,50]$ meV, and the three different lifetimes $\tau_{r}\approx 100/10/1$ ps. The result is displayed in
Fig.~\ref{fig:extracteff_map}. It shows again that the dot-contact coupling value dominates at all lifetimes. For $\tau_{r}\approx 100$ ps, sub-unit photocarrier extraction efficiency occurs only for very weak dot-contact couplings, while at $\tau_{r}\approx 1$ ps, complete extraction may only be achieved for large values of both inter-dot and dot-contact coupling.

\section{Conclusions}

In conclusion, we introduced and applied a customized NEGF simulation framework to provide a direct relation between the local quantum dot array configuration in terms of system size as well as inter-dot and dot-contact coupling strength, and the global photovoltaic device performance of a finite and selectively contacted QD array, mimicking a 1D QD solid. For long carrier lifetime $\tau_{r}\gg 100$ ps and moderate contact coupling $V_{B}<0.2$ eV, the dominant effects result from the dependence of the joint density of states on the inter-dot coupling and the system size in terms of spectral shape, band-width and effective gap. At large contact coupling $V_{B}>0.2$ eV, localized surface states form and have a sizable impact on the optical spectra.  For $\tau_{r}< 100$ ps, where the recombination rates start to compete with the escape rates, the dot-contact coupling becomes the dominant factor for carrier extraction. At an effective carrier lifetime of 1 ps, values of contact coupling $V_{B}<0.1$ eV start to limit carrier extraction at any inter-dot coupling strength. 

The analysis, though performed here for a specific and simplified QD system, can be extended to more complex descriptions of the local electronic structure and to any kind of disorder affecting the charge carrier transport and, hence, the extraction efficiency.

\bibliographystyle{aipnum4-1}

\begin{thebibliography}{36}%
\makeatletter
\providecommand \@ifxundefined [1]{%
 \@ifx{#1\undefined}
}%
\providecommand \@ifnum [1]{%
 \ifnum #1\expandafter \@firstoftwo
 \else \expandafter \@secondoftwo
 \fi
}%
\providecommand \@ifx [1]{%
 \ifx #1\expandafter \@firstoftwo
 \else \expandafter \@secondoftwo
 \fi
}%
\providecommand \natexlab [1]{#1}%
\providecommand \enquote  [1]{``#1''}%
\providecommand \bibnamefont  [1]{#1}%
\providecommand \bibfnamefont [1]{#1}%
\providecommand \citenamefont [1]{#1}%
\providecommand \href@noop [0]{\@secondoftwo}%
\providecommand \href [0]{\begingroup \@sanitize@url \@href}%
\providecommand \@href[1]{\@@startlink{#1}\@@href}%
\providecommand \@@href[1]{\endgroup#1\@@endlink}%
\providecommand \@sanitize@url [0]{\catcode `\\12\catcode `\$12\catcode
  `\&12\catcode `\#12\catcode `\^12\catcode `\_12\catcode `\%12\relax}%
\providecommand \@@startlink[1]{}%
\providecommand \@@endlink[0]{}%
\providecommand \url  [0]{\begingroup\@sanitize@url \@url }%
\providecommand \@url [1]{\endgroup\@href {#1}{\urlprefix }}%
\providecommand \urlprefix  [0]{URL }%
\providecommand \Eprint [0]{\href }%
\providecommand \doibase [0]{http://dx.doi.org/}%
\providecommand \selectlanguage [0]{\@gobble}%
\providecommand \bibinfo  [0]{\@secondoftwo}%
\providecommand \bibfield  [0]{\@secondoftwo}%
\providecommand \translation [1]{[#1]}%
\providecommand \BibitemOpen [0]{}%
\providecommand \bibitemStop [0]{}%
\providecommand \bibitemNoStop [0]{.\EOS\space}%
\providecommand \EOS [0]{\spacefactor3000\relax}%
\providecommand \BibitemShut  [1]{\csname bibitem#1\endcsname}%
\let\auto@bib@innerbib\@empty
\bibitem [{\citenamefont {Green}(2000)}]{green:00}%
  \BibitemOpen
  \bibfield  {author} {\bibinfo {author} {\bibfnamefont {M.~A.}\ \bibnamefont
  {Green}},\ }\href {\doibase DOI: 10.1016/S0921-5107(99)00546-2} {\bibfield
  {journal} {\bibinfo  {journal} {Mater. Sci. Eng., B}\
  }\textbf {\bibinfo {volume} {74}},\ \bibinfo {pages} {118 } (\bibinfo {year}
  {2000})}\BibitemShut {NoStop}%
\bibitem [{\citenamefont {Mart\'i}(2004)}]{marti:04}%
  \BibitemOpen
  \bibfield  {author} {\bibinfo {author} {\bibfnamefont {A.}~\bibnamefont
  {Mart\'i}},\ }\href
  {http://wwwzb.fz-juelich.de/contentenrichment/inhaltsverzeichnisse/2012/0750309059.pdf}
  {\emph {\bibinfo {title} {Next generation photovoltaics : high efficiency
  through full spectrum utilization}}},\ Series in
  optics and optoelectronics\ (\bibinfo  {publisher} {IOP},\ \bibinfo
  {address} {Bristol},\ \bibinfo {year} {2004})\ pp.\ \bibinfo {pages} {XI, 332
  S.}\BibitemShut {Stop}%
\bibitem [{\citenamefont {Tsakalakos}(2008)}]{tsakalakos:08}%
  \BibitemOpen
  \bibfield  {author} {\bibinfo {author} {\bibfnamefont {L.}~\bibnamefont
  {Tsakalakos}},\ }\href {\doibase DOI: 10.1016/j.mser.2008.06.002} {\bibfield
  {journal} {\bibinfo  {journal} {Mater. Sci. Eng., R}\ }\textbf {\bibinfo {volume} {62}},\ \bibinfo {pages} {175 }
  (\bibinfo {year} {2008})}\BibitemShut {NoStop}%
\bibitem [{\citenamefont {Aroutiounian}\ \emph {et~al.}(2001)\citenamefont
  {Aroutiounian}, \citenamefont {Petrosyan}, \citenamefont {Khachatryan},\ and\
  \citenamefont {Touryan}}]{Aroutiounian2001}%
  \BibitemOpen
  \bibfield  {author} {\bibinfo {author} {\bibfnamefont {V.}~\bibnamefont
  {Aroutiounian}}, \bibinfo {author} {\bibfnamefont {S.}~\bibnamefont
  {Petrosyan}}, \bibinfo {author} {\bibfnamefont {A.}~\bibnamefont
  {Khachatryan}}, \ and\ \bibinfo {author} {\bibfnamefont {K.}~\bibnamefont
  {Touryan}},\ }\href {\doibase 10.1063/1.1339210} {\bibfield  {journal}
  {\bibinfo  {journal} {J. Appl. Phys.}\ }\textbf {\bibinfo {volume} {89}},\
  \bibinfo {pages} {2268} (\bibinfo {year} {2001})}\BibitemShut {NoStop}%
\bibitem [{\citenamefont {Nozik}(2002)}]{nozik:02}%
  \BibitemOpen
  \bibfield  {author} {\bibinfo {author} {\bibfnamefont {A.}~\bibnamefont
  {Nozik}},\ }\href@noop {} {\bibfield  {journal} {\bibinfo  {journal} {Physica
  E}\ }\textbf {\bibinfo {volume} {14}},\ \bibinfo {pages} {115} (\bibinfo
  {year} {2002})}\BibitemShut {NoStop}%
\bibitem [{\citenamefont {Raffaelle}\ \emph {et~al.}(2002)\citenamefont
  {Raffaelle}, \citenamefont {Castro}, \citenamefont {Hepp},\ and\
  \citenamefont {Bailey}}]{Raffaelle2002}%
  \BibitemOpen
  \bibfield  {author} {\bibinfo {author} {\bibfnamefont {R.~P.}\ \bibnamefont
  {Raffaelle}}, \bibinfo {author} {\bibfnamefont {S.~L.}\ \bibnamefont
  {Castro}}, \bibinfo {author} {\bibfnamefont {A.~F.}\ \bibnamefont {Hepp}}, \
  and\ \bibinfo {author} {\bibfnamefont {S.~G.}\ \bibnamefont {Bailey}},\
  }\href {\doibase 10.1002/pip.452} {\bibfield  {journal} {\bibinfo  {journal}
  {Prog. Photovolt: Res. Appl.}\ }\textbf {\bibinfo
  {volume} {10}},\ \bibinfo {pages} {433} (\bibinfo {year} {2002})}\BibitemShut
  {NoStop}%
\bibitem [{\citenamefont {Marti}\ \emph {et~al.}(2006)\citenamefont {Mart\'i},
  \citenamefont {L\'opez}, \citenamefont {Antol\'in}, \citenamefont {C\'anovas},
  \citenamefont {Stanley}, \citenamefont {Farmer}, \citenamefont {Cuadra},\
  and\ \citenamefont {Luque}}]{marti:06_tsf}%
  \BibitemOpen
  \bibfield  {author} {\bibinfo {author} {\bibfnamefont {A.}~\bibnamefont
  {Mart\'i}}, \bibinfo {author} {\bibfnamefont {N.}~\bibnamefont {L\'opez}},
  \bibinfo {author} {\bibfnamefont {E.}~\bibnamefont {Antol\'in}}, \bibinfo
  {author} {\bibfnamefont {E.}~\bibnamefont {C\'anovas}}, \bibinfo {author}
  {\bibfnamefont {C.}~\bibnamefont {Stanley}}, \bibinfo {author} {\bibfnamefont
  {C.}~\bibnamefont {Farmer}}, \bibinfo {author} {\bibfnamefont
  {L.}~\bibnamefont {Cuadra}}, \ and\ \bibinfo {author} {\bibfnamefont
  {A.}~\bibnamefont {Luque}},\ }\href {\doibase 10.1016/j.tsf.2005.12.122}
  {\bibfield  {journal} {\bibinfo  {journal} {Thin Solid Films}\ }\textbf
  {\bibinfo {volume} {511 - 512}},\ \bibinfo {pages} {638 – 644} (\bibinfo
  {year} {2006})}\BibitemShut {NoStop}%
\bibitem [{\citenamefont {Oshima}, \citenamefont {Takata},\ and\ \citenamefont
  {Okada}(2008)}]{Oshima2008a}%
  \BibitemOpen
  \bibfield  {author} {\bibinfo {author} {\bibfnamefont {R.}~\bibnamefont
  {Oshima}}, \bibinfo {author} {\bibfnamefont {A.}~\bibnamefont {Takata}}, \
  and\ \bibinfo {author} {\bibfnamefont {Y.}~\bibnamefont {Okada}},\ }\href
  {\doibase 10.1063/1.2973398} {\bibfield  {journal} {\bibinfo  {journal}
  {Appl. Phys. Lett.}\ }\textbf {\bibinfo {volume} {93}},\ \bibinfo {pages}
  {083111} (\bibinfo {year} {2008})}\BibitemShut {NoStop}%
\bibitem [{\citenamefont {Conibeer}\ \emph {et~al.}(2008)\citenamefont
  {Conibeer}, \citenamefont {Green}, \citenamefont {Cho}, \citenamefont
  {K\"{o}nig}, \citenamefont {Cho}, \citenamefont {Fangsuwannarak},
  \citenamefont {Scardera}, \citenamefont {Pink}, \citenamefont {Huang},
  \citenamefont {Puzzer}, \citenamefont {Huang}, \citenamefont {Song},
  \citenamefont {Flynn}, \citenamefont {Park}, \citenamefont {Hao},\ and\
  \citenamefont {Mansfield}}]{Conibeer2008}%
  \BibitemOpen
  \bibfield  {author} {\bibinfo {author} {\bibfnamefont {G.}~\bibnamefont
  {Conibeer}}, \bibinfo {author} {\bibfnamefont {M.}~\bibnamefont {Green}},
  \bibinfo {author} {\bibfnamefont {E.-C.}\ \bibnamefont {Cho}}, \bibinfo
  {author} {\bibfnamefont {D.}~\bibnamefont {K\"{o}nig}}, \bibinfo {author}
  {\bibfnamefont {Y.-H.}\ \bibnamefont {Cho}}, \bibinfo {author} {\bibfnamefont
  {T.}~\bibnamefont {Fangsuwannarak}}, \bibinfo {author} {\bibfnamefont
  {G.}~\bibnamefont {Scardera}}, \bibinfo {author} {\bibfnamefont
  {E.}~\bibnamefont {Pink}}, \bibinfo {author} {\bibfnamefont {Y.}~\bibnamefont
  {Huang}}, \bibinfo {author} {\bibfnamefont {T.}~\bibnamefont {Puzzer}},
  \bibinfo {author} {\bibfnamefont {S.}~\bibnamefont {Huang}}, \bibinfo
  {author} {\bibfnamefont {D.}~\bibnamefont {Song}}, \bibinfo {author}
  {\bibfnamefont {C.}~\bibnamefont {Flynn}}, \bibinfo {author} {\bibfnamefont
  {S.}~\bibnamefont {Park}}, \bibinfo {author} {\bibfnamefont {X.}~\bibnamefont
  {Hao}}, \ and\ \bibinfo {author} {\bibfnamefont {D.}~\bibnamefont
  {Mansfield}},\ }\href {\doibase 10.1016/j.tsf.2007.12.096} {\bibfield
  {journal} {\bibinfo  {journal} {Thin Solid Films}\ }\textbf {\bibinfo
  {volume} {516}},\ \bibinfo {pages} {6748} (\bibinfo {year}
  {2008})}\BibitemShut {NoStop}%
\bibitem [{\citenamefont {Aroutiounian}, \citenamefont {Petrosyan},\ and\
  \citenamefont {Khachatryan}(2005)}]{Aroutiounian2005}%
  \BibitemOpen
  \bibfield  {author} {\bibinfo {author} {\bibfnamefont {V.}~\bibnamefont
  {Aroutiounian}}, \bibinfo {author} {\bibfnamefont {S.}~\bibnamefont
  {Petrosyan}}, \ and\ \bibinfo {author} {\bibfnamefont {A.}~\bibnamefont
  {Khachatryan}},\ }\href {\doibase 10.1016/j.solmat.2005.02.011} {\bibfield
  {journal} {\bibinfo  {journal} {Sol. Energy Mater. Sol. Cells}\
  }\textbf {\bibinfo {volume} {89}},\ \bibinfo {pages} {165} (\bibinfo {year}
  {2005})}\BibitemShut {NoStop}%
\bibitem [{\citenamefont {Shao}\ \emph {et~al.}(2007)\citenamefont {Shao},
  \citenamefont {Balandin}, \citenamefont {Fedoseyev},\ and\ \citenamefont
  {Turowski}}]{Shao2007}%
  \BibitemOpen
  \bibfield  {author} {\bibinfo {author} {\bibfnamefont {Q.}~\bibnamefont
  {Shao}}, \bibinfo {author} {\bibfnamefont {A.~A.}\ \bibnamefont {Balandin}},
  \bibinfo {author} {\bibfnamefont {A.~I.}\ \bibnamefont {Fedoseyev}}, \ and\
  \bibinfo {author} {\bibfnamefont {M.}~\bibnamefont {Turowski}},\ }\href
  {\doibase 10.1063/1.2799172} {\bibfield  {journal} {\bibinfo  {journal}
  {Appl. Phys. Lett.}\ }\textbf {\bibinfo {volume} {91}},\ \bibinfo {pages}
  {163503} (\bibinfo {year} {2007})}\BibitemShut {NoStop}%
\bibitem [{\citenamefont {Gioannini}\ \emph {et~al.}(2013)\citenamefont
  {Gioannini}, \citenamefont {Cedola}, \citenamefont {Santo}, \citenamefont
  {Bertazzi},\ and\ \citenamefont {Cappelluti}}]{Gioannini2013}%
  \BibitemOpen
  \bibfield  {author} {\bibinfo {author} {\bibfnamefont {M.}~\bibnamefont
  {Gioannini}}, \bibinfo {author} {\bibfnamefont {A.~P.}\ \bibnamefont
  {Cedola}}, \bibinfo {author} {\bibfnamefont {N.~D.}\ \bibnamefont {Santo}},
  \bibinfo {author} {\bibfnamefont {F.}~\bibnamefont {Bertazzi}}, \ and\
  \bibinfo {author} {\bibfnamefont {F.}~\bibnamefont {Cappelluti}},\ }\href
  {\doibase 10.1109/JPHOTOV.2013.2270345} {\bibfield  {journal} {\bibinfo
  {journal} {IEEE J. Photovolt.}\ }\textbf {\bibinfo {volume} {3}},\
  \bibinfo {pages} {1271} (\bibinfo {year} {2013})}\BibitemShut {NoStop}%
\bibitem [{\citenamefont {Aeberhard}(2011{\natexlab{a}})}]{ae:jcel_11}%
  \BibitemOpen
  \bibfield  {author} {\bibinfo {author} {\bibfnamefont {U.}~\bibnamefont
  {Aeberhard}},\ }\href {http://dx.doi.org/10.1007/s10825-011-0375-6}
  {\bibfield  {journal} {\bibinfo  {journal} {J. Comput. Electron.}\ }\textbf
  {\bibinfo {volume} {10}},\ \bibinfo {pages} {394} (\bibinfo {year}
  {2011})}\BibitemShut {NoStop}%
\bibitem [{\citenamefont {Aeberhard}(2012)}]{ae:oqel_12}%
  \BibitemOpen
  \bibfield  {author} {\bibinfo {author} {\bibfnamefont {U.}~\bibnamefont
  {Aeberhard}},\ }\href {\doibase 10.1007/s11082-011-9529-9} {\bibfield
  {journal} {\bibinfo  {journal} {Opt. Quantum. Electron.}\ }\textbf {\bibinfo
  {volume} {44}},\ \bibinfo {pages} {133} (\bibinfo {year} {2012})}\BibitemShut
  {NoStop}%
\bibitem [{\citenamefont {Vukmirovic}\ \emph {et~al.}(2007)\citenamefont
  {Vukmirovic}, \citenamefont {Ikonic}, \citenamefont {Indjin},\ and\
  \citenamefont {Harrison}}]{vukmirovic:07}%
  \BibitemOpen
  \bibfield  {author} {\bibinfo {author} {\bibfnamefont {N.}~\bibnamefont
  {Vukmirovic}}, \bibinfo {author} {\bibfnamefont {Z.}~\bibnamefont {Ikonic}},
  \bibinfo {author} {\bibfnamefont {D.}~\bibnamefont {Indjin}}, \ and\ \bibinfo
  {author} {\bibfnamefont {P.}~\bibnamefont {Harrison}},\ }\href {\doibase
  10.1103/PhysRevB.76.245313} {\bibfield  {journal} {\bibinfo  {journal} {Phys.
  Rev. B}\ }\textbf {\bibinfo {volume} {76}},\ \bibinfo {pages} {245313}
  (\bibinfo {year} {2007})}\BibitemShut {NoStop}%
\bibitem [{\citenamefont {Wacker}(2002)}]{wacker:02}%
  \BibitemOpen
  \bibfield  {author} {\bibinfo {author} {\bibfnamefont {A.}~\bibnamefont
  {Wacker}},\ }\href@noop {} {\bibfield  {journal} {\bibinfo  {journal} {Phys.
  Rep.}\ }\textbf {\bibinfo {volume} {357}},\ \bibinfo {pages} {1} (\bibinfo
  {year} {2002})}\BibitemShut {NoStop}%
\bibitem [{\citenamefont {Keldysh}(1965)}]{keldysh:65}%
  \BibitemOpen
  \bibfield  {author} {\bibinfo {author} {\bibfnamefont {L.}~\bibnamefont
  {Keldysh}},\ }\href@noop {} {\bibfield  {journal} {\bibinfo  {journal} {Sov.
  Phys. JETP}\ }\textbf {\bibinfo {volume} {20}},\ \bibinfo {pages} {1018}
  (\bibinfo {year} {1965})}\BibitemShut {NoStop}%
\bibitem [{\citenamefont {Frederiksen}(2004)}]{frederiksen:04}%
  \BibitemOpen
  \bibfield  {author} {\bibinfo {author} {\bibfnamefont {T.}~\bibnamefont
  {Frederiksen}},\ }\emph {\bibinfo {title} {Inelastic electron transport in
  nanosystems}},\ \href@noop {} {Master's thesis},\ \bibinfo  {school}
  {Technical University of Denmark} (\bibinfo {year} {2004})\BibitemShut
  {NoStop}%
\bibitem [{\citenamefont {Aeberhard}\ \emph {et~al.}(2012)\citenamefont
  {Aeberhard}, \citenamefont {Vaxenburg}, \citenamefont {Lifshitz},\ and\
  \citenamefont {Tomic}}]{ae:pccp_12}%
  \BibitemOpen
  \bibfield  {author} {\bibinfo {author} {\bibfnamefont {U.}~\bibnamefont
  {Aeberhard}}, \bibinfo {author} {\bibfnamefont {R.}~\bibnamefont
  {Vaxenburg}}, \bibinfo {author} {\bibfnamefont {E.}~\bibnamefont {Lifshitz}},
  \ and\ \bibinfo {author} {\bibfnamefont {S.}~\bibnamefont {Tomic}},\ }\href
  {\doibase 10.1039/C2CP42213A} {\bibfield  {journal} {\bibinfo  {journal}
  {Phys. Chem. Chem. Phys.}\ }\textbf {\bibinfo {volume} {14}},\ \bibinfo
  {pages} {16223} (\bibinfo {year} {2012})}\BibitemShut {NoStop}%
\bibitem [{\citenamefont {Bartnik}\ \emph {et~al.}(2007)\citenamefont
  {Bartnik}, \citenamefont {Wise}, \citenamefont {Kigel},\ and\ \citenamefont
  {Lifshitz}}]{bartnik:07}%
  \BibitemOpen
  \bibfield  {author} {\bibinfo {author} {\bibfnamefont {A.~C.}\ \bibnamefont
  {Bartnik}}, \bibinfo {author} {\bibfnamefont {F.~W.}\ \bibnamefont {Wise}},
  \bibinfo {author} {\bibfnamefont {A.}~\bibnamefont {Kigel}}, \ and\ \bibinfo
  {author} {\bibfnamefont {E.}~\bibnamefont {Lifshitz}},\ }\href {\doibase
  10.1103/PhysRevB.75.245424} {\bibfield  {journal} {\bibinfo  {journal} {Phys.
  Rev. B}\ }\textbf {\bibinfo {volume} {75}},\ \bibinfo {pages} {245424}
  (\bibinfo {year} {2007})}\BibitemShut {NoStop}%
\bibitem [{\citenamefont {Vaxenburg}\ and\ \citenamefont
  {Lifshitz}(2012)}]{vaxenburg:12}%
  \BibitemOpen
  \bibfield  {author} {\bibinfo {author} {\bibfnamefont {R.}~\bibnamefont
  {Vaxenburg}}\ and\ \bibinfo {author} {\bibfnamefont {E.}~\bibnamefont
  {Lifshitz}},\ }\href {\doibase 10.1103/PhysRevB.85.075304} {\bibfield
  {journal} {\bibinfo  {journal} {Phys. Rev. B}\ }\textbf {\bibinfo {volume}
  {85}},\ \bibinfo {pages} {075304} (\bibinfo {year} {2012})}\BibitemShut
  {NoStop}%
\bibitem [{\citenamefont {Jiang}\ and\ \citenamefont {Green}(2006)}]{jiang:06}%
  \BibitemOpen
  \bibfield  {author} {\bibinfo {author} {\bibfnamefont {C.-W.}\ \bibnamefont
  {Jiang}}\ and\ \bibinfo {author} {\bibfnamefont {M.~A.}\ \bibnamefont
  {Green}},\ }\href {\doibase 10.1063/1.2203394} {\bibfield  {journal}
  {\bibinfo  {journal} {J. Appl. Phys.}\ }\textbf {\bibinfo {volume} {99}},\
  \bibinfo {eid} {114902} (\bibinfo {year} {2006})}\BibitemShut {NoStop}%
\bibitem [{\citenamefont {Tomic}, \citenamefont {Jones},\ and\ \citenamefont
  {Harrison}(2008)}]{tomic:08}%
  \BibitemOpen
  \bibfield  {author} {\bibinfo {author} {\bibfnamefont {S.}~\bibnamefont
  {Tomic}}, \bibinfo {author} {\bibfnamefont {T.~S.}\ \bibnamefont {Jones}}, \
  and\ \bibinfo {author} {\bibfnamefont {N.~M.}\ \bibnamefont {Harrison}},\
  }\href {\doibase 10.1063/1.3058716} {\bibfield  {journal} {\bibinfo
  {journal} {Appl. Phys. Lett.}\ }\textbf {\bibinfo {volume} {93}},\ \bibinfo
  {pages} {263105} (\bibinfo {year} {2008})}\BibitemShut {NoStop}%
\bibitem [{\citenamefont {Lake}\ \emph {et~al.}(1997)\citenamefont {Lake},
  \citenamefont {Klimeck}, \citenamefont {Bowen},\ and\ \citenamefont
  {Jovanovic}}]{lake:97}%
  \BibitemOpen
  \bibfield  {author} {\bibinfo {author} {\bibfnamefont {R.}~\bibnamefont
  {Lake}}, \bibinfo {author} {\bibfnamefont {G.}~\bibnamefont {Klimeck}},
  \bibinfo {author} {\bibfnamefont {R.}~\bibnamefont {Bowen}}, \ and\ \bibinfo
  {author} {\bibfnamefont {D.}~\bibnamefont {Jovanovic}},\ }\href@noop {}
  {\bibfield  {journal} {\bibinfo  {journal} {J. Appl. Phys.}\ }\textbf
  {\bibinfo {volume} {81}},\ \bibinfo {pages} {7845} (\bibinfo {year}
  {1997})}\BibitemShut {NoStop}%
\bibitem [{Note1()}]{Note1}%
  \BibitemOpen
  \bibinfo {note} {A perturbative approach might not always apply in QD array
  systems. In some cases of strong electron-phonon interaction, a polaronic
  picture might be more appropriate. In the present work, this description is
  merely used as a phenomenological, but still current conserving broadening
  mechanism}\BibitemShut {NoStop}%
\bibitem [{\citenamefont {Aeberhard}(2014{\natexlab{a}})}]{ae:prb89_14}%
  \BibitemOpen
  \bibfield  {author} {\bibinfo {author} {\bibfnamefont {U.}~\bibnamefont
  {Aeberhard}},\ }\href@noop {} {\bibfield  {journal} {\bibinfo  {journal}
  {Phys. Rev. B}\ }\textbf {\bibinfo {volume} {89}},\ \bibinfo {pages} {115303}
  (\bibinfo {year} {2014}{\natexlab{a}})}\BibitemShut {NoStop}%
\bibitem [{\citenamefont {Aeberhard}(2011{\natexlab{b}})}]{ae:prb_11}%
  \BibitemOpen
  \bibfield  {author} {\bibinfo {author} {\bibfnamefont {U.}~\bibnamefont
  {Aeberhard}},\ }\href {\doibase 10.1103/PhysRevB.84.035454} {\bibfield
  {journal} {\bibinfo  {journal} {Phys. Rev. B}\ }\textbf {\bibinfo {volume}
  {84}},\ \bibinfo {pages} {035454} (\bibinfo {year}
  {2011}{\natexlab{b}})}\BibitemShut {NoStop}%
\bibitem [{\citenamefont {Pourfath}\ and\ \citenamefont
  {Kosina}(2009)}]{pourfath:09}%
  \BibitemOpen
  \bibfield  {author} {\bibinfo {author} {\bibfnamefont {M.}~\bibnamefont
  {Pourfath}}\ and\ \bibinfo {author} {\bibfnamefont {H.}~\bibnamefont
  {Kosina}},\ }\href@noop {} {\bibfield  {journal} {\bibinfo  {journal}
  {J. Comput. Electron.}\ } (\bibinfo {year}
  {2009})}\BibitemShut {NoStop}%
\bibitem [{\citenamefont {Aeberhard}(2014{\natexlab{b}})}]{ae:jpe_14}%
  \BibitemOpen
  \bibfield  {author} {\bibinfo {author} {\bibfnamefont {U.}~\bibnamefont
  {Aeberhard}},\ }\href {\doibase 10.1117/1.JPE.4.042099} {\bibfield  {journal}
  {\bibinfo  {journal} {J. Photon. Energy}\ }\textbf {\bibinfo {volume} {4}},\
  \bibinfo {pages} {042099} (\bibinfo {year} {2014}{\natexlab{b}})}\BibitemShut
  {NoStop}%
\bibitem [{\citenamefont {Lake}\ and\ \citenamefont {Pandey}(2006)}]{lake:06}%
  \BibitemOpen
  \bibfield  {author} {\bibinfo {author} {\bibfnamefont {R.~K.}\ \bibnamefont
  {Lake}}\ and\ \bibinfo {author} {\bibfnamefont {R.~R.}\ \bibnamefont
  {Pandey}},\ }in\ \href@noop {} {\emph {\bibinfo {booktitle} {Handbook of
  Semiconductor Nanostructures}}},\ \bibinfo {editor} {edited by\ \bibinfo
  {editor} {\bibfnamefont {A.~A.}\ \bibnamefont {Balandin}}\ and\ \bibinfo
  {editor} {\bibfnamefont {K.~L.}\ \bibnamefont {Wang}}}\ (\bibinfo
  {publisher} {American Scientific Publishers},\ \bibinfo {year}
  {2006})\BibitemShut {NoStop}%
\bibitem [{\citenamefont {Pereira}\ and\ \citenamefont
  {Henneberger}(1998)}]{pereira:98}%
  \BibitemOpen
  \bibfield  {author} {\bibinfo {author} {\bibfnamefont {M.~F.}\ \bibnamefont
  {Pereira}}\ and\ \bibinfo {author} {\bibfnamefont {K.}~\bibnamefont
  {Henneberger}},\ }\href@noop {} {\bibfield  {journal} {\bibinfo  {journal}
  {Phys. Rev. B}\ }\textbf {\bibinfo {volume} {58}},\ \bibinfo {pages} {2064}
  (\bibinfo {year} {1998})}\BibitemShut {NoStop}%
\bibitem [{\citenamefont {Richter}, \citenamefont {Florian},\ and\
  \citenamefont {Henneberger}(2008)}]{richter:08}%
  \BibitemOpen
  \bibfield  {author} {\bibinfo {author} {\bibfnamefont {F.}~\bibnamefont
  {Richter}}, \bibinfo {author} {\bibfnamefont {M.}~\bibnamefont {Florian}}, \
  and\ \bibinfo {author} {\bibfnamefont {K.}~\bibnamefont {Henneberger}},\
  }\href {\doibase 10.1103/PhysRevB.78.205114} {\bibfield  {journal} {\bibinfo
  {journal} {Phys. Rev. B}\ }\textbf {\bibinfo {volume} {78}},\ \bibinfo {eid}
  {205114} (\bibinfo {year} {2008})}\BibitemShut {NoStop}%
\bibitem [{\citenamefont {W\"urfel}(1982)}]{wuerfel:82}%
  \BibitemOpen
  \bibfield  {author} {\bibinfo {author} {\bibfnamefont {P.}~\bibnamefont
  {W\"urfel}},\ }\href@noop {} {\bibfield  {journal} {\bibinfo  {journal} {J.
  Phys. C: Solid State Phys.}\ }\textbf {\bibinfo {volume} {15}},\ \bibinfo
  {pages} {3967} (\bibinfo {year} {1982})}\BibitemShut {NoStop}%
\bibitem [{Note2()}]{Note2}%
  \BibitemOpen
  \bibinfo {note} {Strictly, this relation between absorption at zero bias and
  emission under finite QFLS is only valid if neither JDOS nor the optical
  matrix elements $\protect \mathcal {M}^{e\gamma }$ are modified at finite
  bias voltage, e.g., due to a change in orbital overlap or local band
  bending.}\BibitemShut {Stop}%
\bibitem [{Note3()}]{Note3}%
  \BibitemOpen
  \bibinfo {note} {In order to reproduce lifetimes of recombination processes
  that compete with carrier extraction, the radiative recombination is enhanced
  by a corresponding factor in the self-energy of the spontaneous
  emission}\BibitemShut {NoStop}%
\bibitem [{Note4()}]{Note4}%
  \BibitemOpen
  \bibinfo {note} {The upper bound to $V_{B}$ is chosen in order to exclude
  spurious effects of surface states.}\BibitemShut {Stop}%
\end{thebibliography}
%

\end{document}